\documentclass[aps, twocolumn]{revtex4-2}
\usepackage [dvips]{graphicx}
\usepackage{amsmath}
\usepackage{amssymb}
\usepackage{color}
\usepackage{mathrsfs}
\usepackage{appendix}
\usepackage{ulem}\usepackage{stackengine}
\usepackage[margin=1in]{geometry}
\usepackage{hyperref}

\newcommand{\bomega}{\boldsymbol{\omega}}
\newcommand{\bGamma}{\boldsymbol{\Gamma}}

\newcommand{\dd}{\text{d}}

\newcommand{\ee}{\text{e}}

\newcommand{\bx}{\boldsymbol x}
\newcommand{\br}{\boldsymbol r}

\newcommand{\bff}{\boldsymbol f}

\newcommand{\bp}{\boldsymbol p}
\newcommand{\bc}{\boldsymbol c}
\newcommand{\bu}{\boldsymbol u}
\newcommand{\bU}{\text{\bf U}}

\newcommand{\bg}{{\bf g}}

\newcommand{\bv}{\boldsymbol v}

\newcommand{\bnabla}{\boldsymbol{\nabla}}

\newcommand{\bn}{\boldsymbol n}

\newcommand{\bF}{\boldsymbol F}

\usepackage[percent]{overpic}

\begin{document}

\title{The Anomalous Long-Ranged Influence of an Inclusion in Momentum-Conserving Active Fluids}

\author{Thibaut Arnoulx de Pirey}
\affiliation{Department of Physics, Technion-Israel Institute of Technology, Haifa
32000, Israel}
\author{Yariv Kafri}
\affiliation{Department of Physics, Technion-Israel Institute of Technology, Haifa
32000, Israel}
\author{Sriram Ramaswamy}
\affiliation{Centre for Condensed Matter Theory, Department of Physics, Indian Institute of Science, Bangalore 560 012, India}

\begin{abstract}
We show that an inclusion placed inside a dilute Stokesian suspension of microswimmers induces power-law number-density modulations and flows. These take a different form depending on whether the inclusion is held fixed by an external force, for example an optical tweezer, or if it is free. When the inclusion is held in place, the far-field fluid flow is a Stokeslet, while the microswimmer density decays as $1/r^{2+\epsilon}$, with $r$ the distance from the inclusion, and $\epsilon$ an anomalous exponent which depends on the symmetry of the inclusion and varies continuously as a function of a dimensionless number characterizing the relative amplitudes of the convective and diffusive effects. The angular dependence takes a non-trivial form which depends on the same dimensionless number. When the inclusion is free to move, the far-field fluid flow is a stresslet and the microswimmer density decays as $1/r^2$ with a simple angular dependence. These long-range modulations mediate long-range interactions between inclusions that we characterize.

\end{abstract}

\maketitle

\section{Introduction}

Active matter encompasses systems whose individual elements convert energy into directed motion on a microscopic scale~\cite{marchetti2013hydrodynamics,ramaswamy2010mechanics,prost2015active,bechinger2016active,dadhichi2018origins,fodor2018statistical,chate2020dry,tailleur2022active,o2022time}. When the dissipative conversion of energy is coupled to interactions between particles, a wealth of phenomena which is not exhibited by systems in the thermal equilibrium is observed. Similarly, when this breaking of time-reversal symmetry is coupled to interactions with external potentials the resulting behavior is very different than that of equilibrium systems. Importantly, in equilibrium, when interactions are local, the Boltzmann weight implies that the effect of a localized external potential extends beyond its own support only out to a scale of order the correlation length.
In stark contrast, in active systems with local conservation laws, steady-state distributions  are inherently non-local~\cite{solon2015pressure,woillez2020nonlocal,woillez2019activated,o2022time,granek2023inclusions} which leads to long-ranged influences of external potentials. A particularly spectacular experimental manifestation is the response of active systems to asymmetric potentials placed in the middle of a chamber~\cite{galajda2007wall}. One finds that active particles accumulate on one side of the system as a result of a ratchet-like mechanism~\cite{tailleur2009sedimentation}.

Much theoretical progress has been made in understanding the response of active matter to external potentials in {\it dry} active systems. In dry systems momentum is not conserved, so that experimental realizations correspond, for example, to particles moving on a substrate \cite{wolgemuth2002myxobacteria}, vibrating granular grains \cite{narayan2007long,deseigne2010collective}, and more. Significant attention has been given to the particle density in confining potentials \cite{ezhilan2015distribution, wagner2017steady, tailleur2009sedimentation, smith2022exact} and in the vicinity of localized repulsive potentials \cite{kaiser2012capture, kumar2019trapping, dePirey_2023}, showing the generic tendency for active particles to accumulate close to walls and repulsive boundaries. Arguably equally significant is the observation that generic localized potentials (or inclusions) induce a universal long-range modulation of the density field \cite{baek2018generic, granek2020bodies, speck2021vorticity} which decays $\propto {\bf p}\cdot {\bf r}/r^{d}$ in $d$ dimensions, with ${\bf p}$ a vector characterizing the properties of the inclusion and ${\bf r}$ is the distance from it. The behavior is a consequence of the emergence of ratchet currents from the interplay between the breaking of time-reversal symmetry and any asymmetry of the inclusion. The result has far-reaching consequences~\cite{granek2023inclusions}. It implies that two inclusions placed in an active bath experience long-range interactions~\cite{baek2018generic,granek2020bodies,ni2015tunable} and explains the sensitivity of the phase diagram of dry active systems to bulk~\cite{ro2021disorder} and boundary~\cite{dor2022disordered} disorder. In particular, quenched disorder generically leads to long-range correlations \cite{ro2021disorder} in any dilute active system. Moreover, motility-induced-phase-separation \cite{tailleur2008statistical,fily2012athermal,redner2013structure,cates2015motility} is destroyed by bulk disorder in dimensions $d<4$, and by boundary disorder in dimensions $d<3$.

\onecolumngrid
\begin{center}
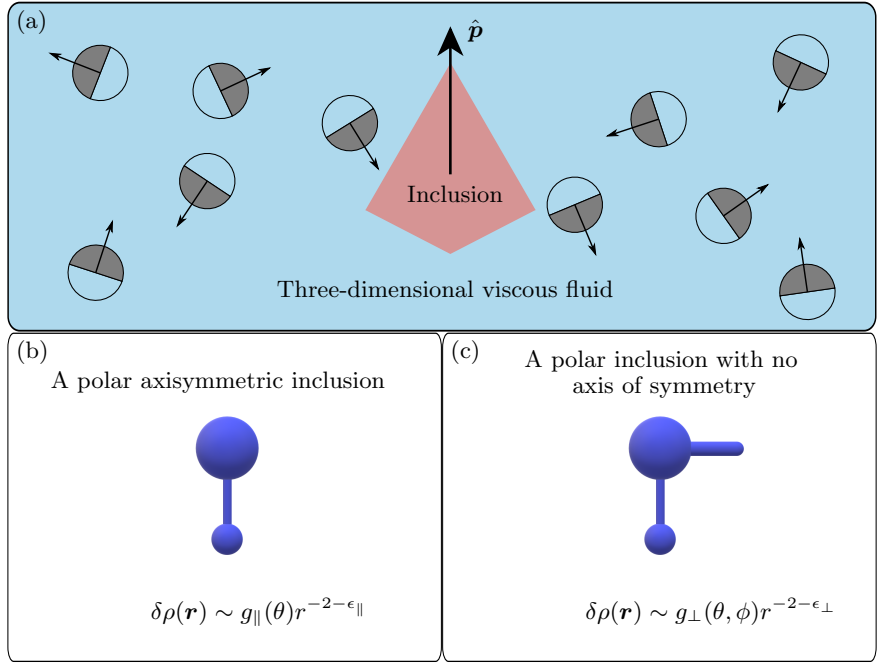
\begin{figure}
\begin{overpic}[scale=.15]{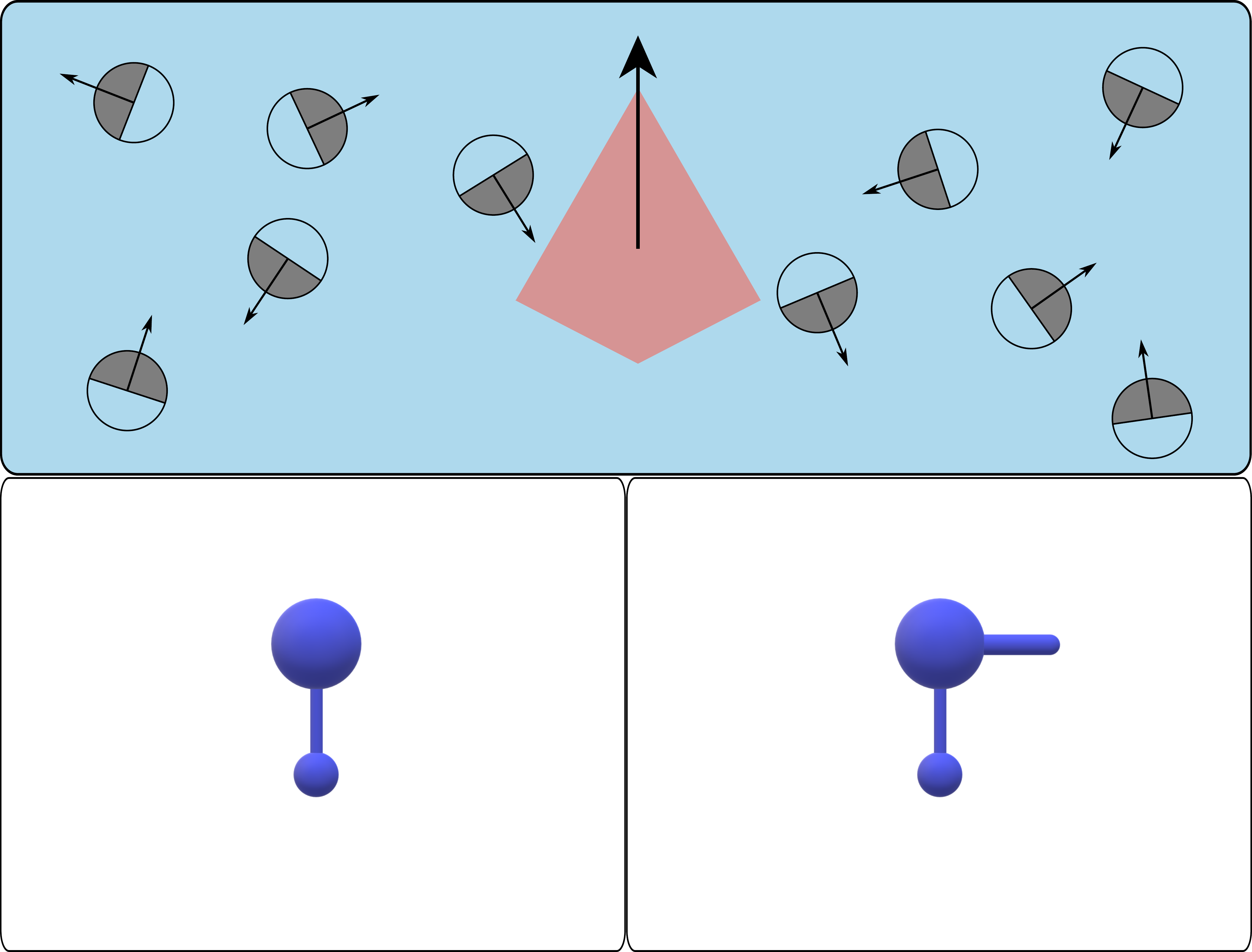}
\put(31,42){Three-dimensional viscous fluid}
\put(46, 53){Inclusion}
\put(53, 72){$\hat{\bp}$}
\put(5, 31.5){A polar axisymmetric inclusion}
\put(59.5, 34){A polar inclusion with no}
\put(65, 31){axis of symmetry}
\put(16.5, 5){$\delta\rho(\br) \sim g_\parallel(\theta) r^{-2-\epsilon_\parallel}$}
\put(66.5, 5){$\delta\rho(\br) \sim g_\perp(\theta,\phi)r^{-2-\epsilon_\perp}$}

\put(1, 73){(a)}
\put(1, 35){(b)}
\put(51, 35){(c)}
\end{overpic}
\caption{Panel (a) is a sketch of the system under consideration: Self-propelled particles swimming in a 3-dimensional Newtonian viscous fluid in the presence of a localized inclusion. The unit vector $\hat{\bp}$ is defined in Eq.~\eqref{eq:pmono} and points in the direction of the average force that must be exerted on the inclusion in order to maintain it fixed. Panels (b, c) illustrate our key finding: a localized inclusion induces a long-range modulation of the density field, whose exponent depends on the symmetries of the inclusion. For fixed polar axisymmetric inclusions, we obtain $\delta\rho(\br) \sim g_\parallel(\theta)r^{-2-\epsilon_\parallel}$ while for those with no axis of symmetry, we get $\delta\rho(\br) \sim g_\perp(\theta,\phi)r^{-2-\epsilon_\perp}$  where $g_\parallel(\theta)$, $g_\perp(\theta,\phi)$, $\epsilon_\parallel$ and $\epsilon_\perp$ are given in Eqs.~(\ref{eq:heur_res}-\ref{eq:heur_res_2bis}).}
\label{fig:cartoon}
\end{figure}
\end{center}
\twocolumngrid

Despite the relevance of dry active matter to experiments, many realizations of active systems, biological or synthetic, comprise particles that self-propel in a viscous fluid. In such systems, termed ``wet'', the conservation of momentum is known to lead to very different behaviors \cite{simha2002hydrodynamic,voituriez2005spontaneous,marchetti2013hydrodynamics,maitra2023two,tiribocchi2015active,baskaran2009statistical}. The dynamics of active particles in wet systems, which in this context are often called microswimmers, in the vicinity of walls and obstacles have been the subject of intense scrutiny \cite{takagi2014hydrodynamic, spagnolie2015geometric, li2014hydrodynamic, elgeti2016microswimmers}. However, the response to a localized inclusion has, to the best of our knowledge, remained unexplored. In this work, we investigate the long-range effect of a localized inclusion by considering a dilute suspension of swimmers propelling in a three-dimensional viscous fluid, as depicted in Fig.~\ref{fig:cartoon}. The presence of the ambient fluid mediates interactions between the particles, which are long-range due to momentum conservation \cite{felderhof1977hydrodynamic}. Direct, non-hydrodynamic, interactions between the swimmers are neglected but are taken into account between the swimmers and the obstacle as a short-ranged force field. As we show, the coupling to fluid flow can qualitatively alter the nature of the long-range effect, and in ways not revealed by mere power-counting.

We identify three cases of interest, corresponding to three different large-scale behaviors of the density field of the swimmers, depending on whether the inclusion is freely moving in the fluid or if it is held fixed by an external force, for instance by optical tweezers, and depending on the internal symmetries of the inclusion. Our results are largely independent of the intrinsic complexity of the near-obstacle swimming motion. When the obstacle is freely moving, driven by the interactions with the swimming particles, hydrodynamic interactions have little impact on the far-field behavior of the density field, and the behavior of the dry case survives with modulations of the density field decaying as $1/r^2$. However, we predict a very different response when the obstacle is held fixed by an external force. In this case, the decay exponent depends on the symmetries of the object and on a Péclet number, called $\lambda$ in the following, that compares the relative amplitude of hydrodynamic to diffusive effects and whose mathematical expression is given in Eq.~\eqref{eq:lambda}. We find that obstacles with a polarity that also defines an axis of (possibly discrete) rotational symmetry induce density modulations decaying as $1/r^{2+\epsilon_\parallel}$ with $\epsilon_\parallel > 0$ while less symmetric obstacles induce density modulations decaying as $1/r^{2+\epsilon_\perp}$ with $\epsilon_\perp < 0$. Lastly, obstacles with no polarity induce, as in the dry case, shorter-ranged density modulations. Notably, we expect density modulations induced by spherical obstacles to decay faster than a power-law.

We begin in Sec.~\ref{sec:heuristic} by presenting a heuristic approach to the effect of hydrodynamic interactions on the behavior of the number density field far away from a localized inclusion. The range of results we obtain are stated at the end of this section. This heuristics is supported by the use of a microscopic model of squirmers that we present in Sec.~\ref{sec:micro} and for which we derive, in a mean-field approximation, the equation obeyed by the steady-state density profile of the swimming particles. We solve this equation in the far-field in Sec.~\ref{sec:asympt}, using an asymptotic expansion of the second kind \cite{goldenfeld2018lectures,barenblatt1996scaling}. We obtain the decay exponent and associated angular dependence of the density field perturbatively in the parameter $\lambda$. An alternative route to these results, based on the renormalization group, is presented in App.~\ref{app:RG}. Finally, before concluding, we build in Sec.~\ref{sec:interactions} on the previous sections to derive the far-field interaction between two inclusions in a bath of swimmers. Throughout, vectors are denoted in bold $\bp$ or in component notation $p^\alpha$ and $\hat{\bp}$ is the unit vector $\hat{\bp}=\bp/|\bp|$.

\section{Heuristic arguments} \label{sec:heuristic}
Before turning to a systematic derivation, we start by presenting the physical picture that underlies the results. It is useful to first consider the dry case. In this case, the localized asymmetric object, through a ratchet effect, acts as a pump on the active particles. Since the active particles diffuse on large scales, the steady-state density $\rho(\br)$ is controlled by the equation $D \partial_\alpha \partial^\alpha \rho(\br)  = - \partial_\alpha C^\alpha(\br)$.
Here $D$ is a diffusion constant, the boundary conditions are $\rho(\br) \to \rho_0$ as $r \equiv |\br| \to\infty$, and $C^\alpha(\br)$ is a current term localized in the vicinity of the obstacle which accounts for near-field effects. Taking $\br = 0$ as the position of the obstacle, it is easy to check that the known far-field behavior, described in the introduction, is captured by this equation as long as $c^\alpha = \int \dd \br \, C^\alpha(\br)$ is finite. The addition of a three-dimensional viscous fluid, because of the long-range nature of hydrodynamic interactions, then modifies the diffusive behavior of the swimmers according to  
\begin{equation} \label{eq:density_field}
D \partial_\alpha \partial^\alpha \rho(\br) - \partial_\alpha \left(\bar{v}^\alpha(\br) \rho(\br)\right) = - \partial_\alpha C^\alpha(\br) \,,
\end{equation}
where $\bar{\bv}(\br)$ is an effective long-ranged convective flow generated by the combined effect of the swimmers and the object. In Sec.~\ref{sec:micro} we show that Eq.~\eqref{eq:density_field} can be derived from a mean-field microscopic model of swimmers. Note that if the obstacle is moving, we assume that it does so on a time scale that is slow enough that the density $\rho(\br)$ can be taken to be in a steady state.

While the microscopic derivation also makes the form of the velocity field $\bar{\bv}(\br)$ explicit, it can be understood intuitively using momentum conservation. Denote by $\bF^i_{{\rm swim} \to\rm{fluid}}$ the force exerted on the fluid by the swimmer labeled by $i$. Since its inertia is negligible, and in the absence of non-hydrodynamic interactions between swimmers, momentum conservation implies that $\bF^i_{{\rm swim} \to\rm{fluid}} = -\bF^i_{{\rm swim} \, \to \rm{obs}}$ where $\bF^i_{{\rm swim} \, \to \rm{obs}}$ is the force exerted by swimmer $i$ on the obstacle. By assumption, the latter is non-zero only for particles in the vicinity of the obstacle. Denote now by $\bF_{\rm{fluid}\to \rm{obs}}$ the force exerted by the fluid on the obstacle. The total force exerted by the combined effect of the swimmers and the obstacle on the fluid, denoted by $\bff$, is therefore
\begin{equation}
    \bff = - \left(\bF_{\rm{fluid}\to \rm{obs}} + \sum_i \bF^i_{{\rm swim} \, \to \rm{obs}}\right) \,.
\end{equation}
In the far-field, this induces a viscous flow, corresponding to a force monopole localized at $\br = 0$ with amplitude $\bff$. It follows that two distinct cases need to be distinguished, depending on whether the obstacle is held fixed externally or not. 

If the obstacle is held fixed by an external force, momentum is injected locally into the system, and $\bff = \bF_{\rm{ext}}$ with $\bF_{\rm{ext}}$ the force exerted by the external observer. Accordingly, the effective flow in Eq.~\eqref{eq:density_field} behaves as a Stokeslet on large scales and we find
\begin{equation} \label{eq:stokesflow}
    \bar{v}^\alpha(\br) \simeq \frac{1}{8\pi\eta}J^{\alpha\beta}(\br) \overline{\bF_{\rm{ext}}} \,,
\end{equation}
where the overline denotes a steady-state average of $\bF_{\rm{ext}}$ which on symmetry grounds is non-zero for a polar obstacle. Here,
\begin{equation}
    J^{\alpha\beta}(\br) = \frac{\delta^{\alpha\beta}}{r} + \frac{r^\alpha r^\beta}{r^3} \,,
\end{equation}
is the fundamental solution of the Stokes equation in the presence of a force monopole. Note that the flow $\bar{\bv}(\br)$ decreases as $r^{-1}$ away from the obstacle. A second case of interest is that of a free obstacle. Here, the total momentum is conserved and $\bff = 0$ so that the leading order far-field effective flow is that of a force dipole 
\begin{equation} \label{eq:dipoleflow}
    \bar{v}^\alpha(\br) \simeq \frac{1}{8\pi\eta}\partial_\gamma J^{\alpha\beta}(\br)Q^{\gamma\beta} \,,
\end{equation}
with $Q^{\gamma\beta}$ the effective average dipole strength. In this case, $ \bar{\bv}(\br)$ decays as $r^{-2}$. 

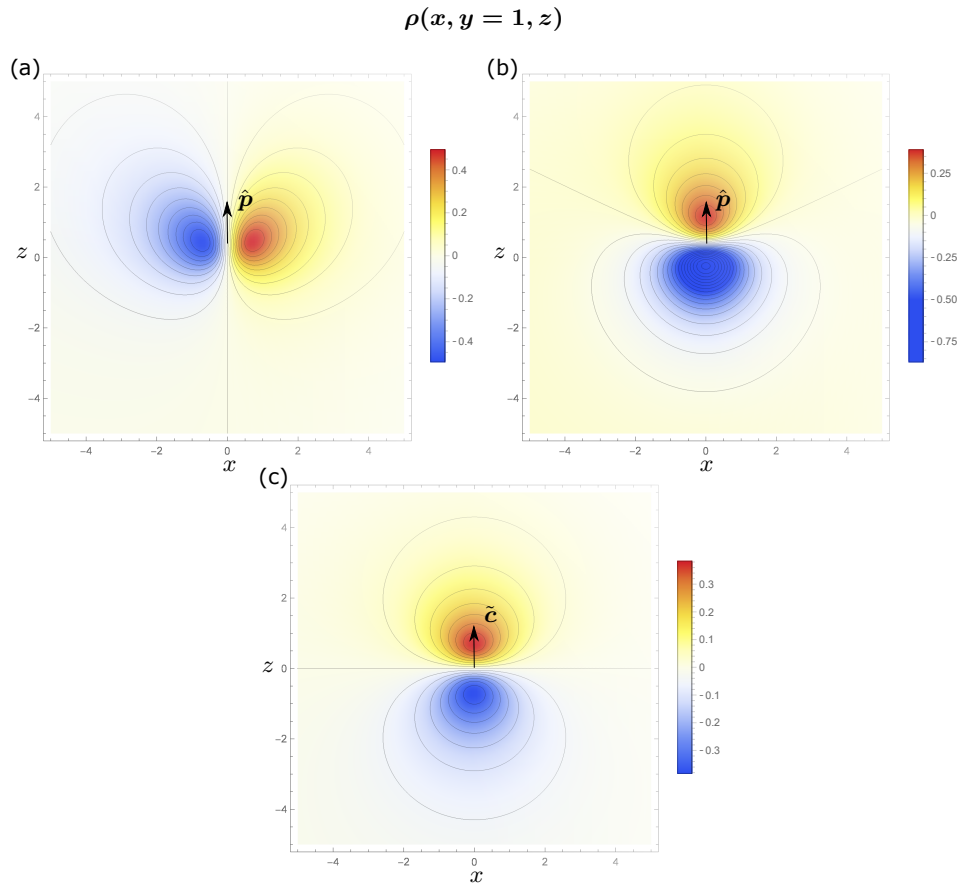
\begin{figure}
\centering

${\boldsymbol{\rho(x,y=1,z)}}$\par\medskip
\vspace{0.2cm}

\begin{overpic}[scale=0.4]{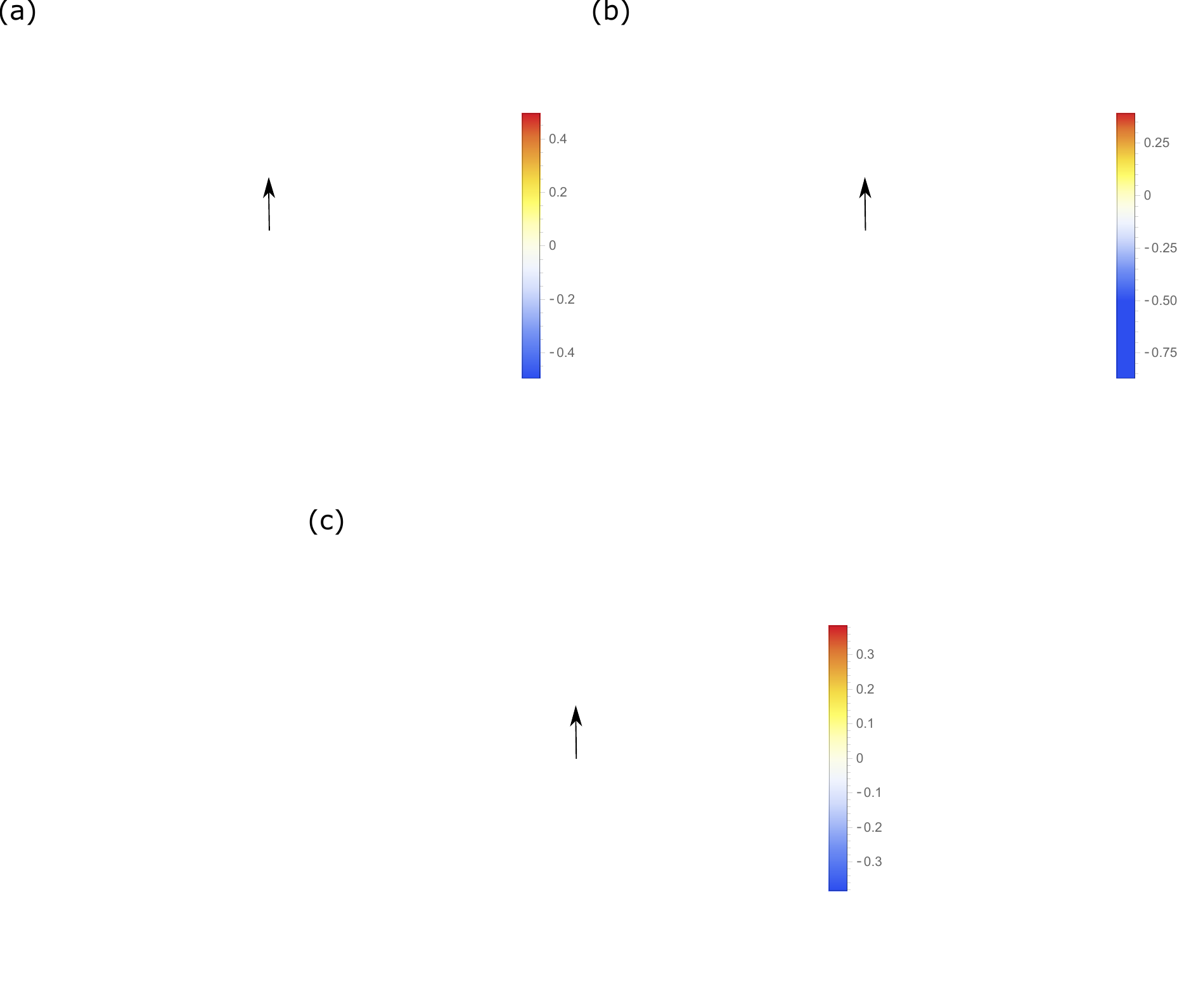}
\put(0.5,64.5){$z$}
\put(22.4,42){$x$}

\put(51,64.5){$z$}
\put(72.8,42){$x$}

\put(26.5,20.7){$z$}
\put(48.5,-1.5){$x$}

\put(24,70){$\hat{\bp}$}
\put(74.5,70){$\hat{\bp}$}
\put(50,26){$\tilde{\bc}$}
\end{overpic}

\caption{\textbf{Far-field density profile} in a two-dimensional section $\rho(x,y=1,z)$, up to a multiplicative constant, for the three different cases: fixed inclusion with no axis of symmetry, fixed polar inclusion with an axis of symmetry and freely-moving inclusion. (a) Fixed inclusion with no axis of symmetry. The vector $\hat{\bp}$ giving the direction of the force monopole is taken along the $z$-axis. The $x$-axis is defined such that the phase $\phi_0$ vanishes in spherical coordinates of axis $(x,y,z)$. (b) Fixed polar inclusion with an axis of symmetry. The vector $\hat{\bp}$ giving the direction of the force monopole is taken along the $z$-axis. In both (a) and (b), we used the second order expansion in $\lambda$ in Eqs.~\eqref{eq:heur_res_bis}-\eqref{eq:heur_res_2bis} and plotted the results taking $\lambda = 1$. (c) Freely-moving polar inclusion. The vector $\tilde{c}^\alpha$ entering Eq.~\eqref{eq:result_freely_moving} is taken along the $z$-axis.}
\label{fig:density_plots}
\end{figure}

As we now argue, the difference in the decay of the velocity field between these two cases results in drastically different behaviors for the density field which, in general, cannot be inferred using simple power counting. This can be understood through the following asymptotic arguments. Denote $\delta\rho(\br) \equiv \rho(\br) - \rho_0$ such that $\delta\rho(\br) \to 0$ as $r \to \infty$. In the far-field, we replace the localized current $C^\alpha(\br)$ by $c^\alpha \delta(\br)$ and the velocity field by $\bar{v}^\alpha(\br) = A r^{-\chi} g^\alpha(\hat{\br})$, with $g^\alpha(\hat{\br})$ controlling the angular dependence. Here $\chi$ is treated as a variable and we keep in mind that $\chi= 1$ corresponds to an externally held obstacle, and $\chi = 2$ to the freely-moving one. The parameter $A$ measures the strength of the hydrodynamic term and can be read from Eq. \eqref{eq:stokesflow} for a fixed obstacle and Eq. \eqref{eq:dipoleflow} for a free obstacle. Since the flow field is incompressible we have
\begin{equation}\label{eq:to_rescale}
    D \Delta \delta\rho - A r^{-\chi} \bg(\hat{\br})  \cdot \bnabla \delta\rho  = - \bc\cdot\bnabla \delta(\br) \;.
\end{equation}
Now, note that if $\chi > 1$ the convection term decays faster at infinity than the diffusive one, rendering the former irrelevant on large length scales. However, both have the same amplitude when $\chi= 1$ indicating that the convection term is marginal in the renormalization group sense and could modify the far-field decay of the density~\footnote{Indeed, when $A = 0$, the equation is left invariant by rescaling space according to $\br \to \br' =  b^{-1} \, \br$ and the density field according to $\delta\rho(\br) \to \delta\rho'(\br') = b^{2}\delta\rho(\br)$. Implementing the same rescaling in Eq.~\ref{eq:to_rescale}, we obtain the following equation
\begin{equation*}
D \Delta' \delta\rho' + A b^{1-\chi}\,\bnabla \cdot \left(r'^{-\chi} \bg(\hat{\br}) \delta\rho' \right) = - {\bf c}\cdot\bnabla' \delta(\br') \,.
\end{equation*}
Here coupling to hydrodynamics is irrelevant, in the sense of the renormalization group, if $\chi > 1$ and is marginal for $\chi = 1$.} With this in mind, we find the following behaviors for fixed and free obstacles embedded in three-dimensional active suspensions. The results are depicted in Fig.~\ref{fig:density_plots} in the three cases of interest that we identify.

\paragraph{Fixed obstacle:} We treat the hydrodynamic coupling using an intermediate asymptotic expansion of the second kind \cite{barenblatt1996scaling} in Sec.~\ref{sec:asympt}, and a renormalization group analysis in Appendix~\ref{app:RG}. We find that the decay of the density field exhibits an anomalous exponent and an angular dependence which depend on the dimensionless parameter $\lambda$ which quantifies the relative amplitude of the diffusive and convective terms,
\begin{equation} \label{eq:lambda}
    \lambda = \frac{\left|\,\overline{\bF_{\rm{ext}}}\,\right|}{8\pi \eta D} \,,
\end{equation}
and on the unit vector,
\begin{equation} \label{eq:pmono}
   \hat{\bp} = \frac{\overline{\bF_{\rm{ext}}}}{\left|\,\overline{\bF_{\rm{ext}}}\,\right|} \,,
\end{equation}
which points along the force monopole. Note that local injection of angular momentum leads to flow fields decaying as $r^{-2}$ which is why, following the reasoning below Eq.~\eqref{eq:to_rescale}, the large scale behavior of the density field is insensitive to the total external torque exerted on the obstacle, if any. A striking feature is that the anomalous exponent and the angular dependence also depend on the symmetry of the obstacle. Our results are expressed as a perturbative expansion in powers of $\lambda$, which is relevant for dilute suspensions where $\lambda$ is small (because the force in the numerator of Eq.~\eqref{eq:lambda} scales as the density of active particles at low density).
\\ For obstacles for which the vector $\hat{\bp}$ defines an axis of (possibly discrete) rotational symmetry, we obtain
\begin{equation}\label{eq:heur_res}
    \delta\rho(\br) \sim \frac{g_\parallel(\theta)}{r^{2 + \epsilon_{\parallel}}} \,\,\,\, \rm{with} \,\,\,\, \epsilon_{\parallel} = \frac{\lambda^2}{3} + O(\lambda^4) \,,
\end{equation}
where $\theta$ is the angle between $\hat{\br}$ and $\hat{\bp}$. The density field, therefore, decays {\it faster} than in the absence of hydrodynamic interactions. The angular dependence is given to order $O(\lambda^2)$ by
\begin{equation}\label{eq:heur_res_bis}
g_\parallel(\theta) = \cos\theta - \frac{\lambda}{4}\left(3  - 5 \cos^2\theta\right) + \frac{3}{4} \lambda^2 \cos^3\theta \,.
\end{equation}
However, for obstacles with no axis of symmetry, the density field also depends on the azimuthal angle $\phi$ of spherical coordinates of axis $\hat{\bp}$ and features a different exponent,
\begin{equation}\label{eq:heur_res_2}
    \delta\rho(\br) \sim \frac{g_\perp(\theta,\phi)}{r^{2 + \epsilon_{\perp}}} \,\,\,\, \rm{with} \,\,\,\, \epsilon_{\perp} = -\frac{\lambda^2}{12} + O(\lambda^4) \,,
\end{equation}
showing that the decay is {\it slower} than in the absence of hydrodynamic interactions. To order $O(\lambda^2)$, the angular dependence is given by
\begin{align}\label{eq:heur_res_2bis}
    g_\perp(\theta,\phi)&  =  \cos(\phi + \phi_0)\sin(\theta) \nonumber \\ & \times \left(1 + \frac{5\lambda}{4}\cos\theta + \frac{3}{4} \lambda^2 \cos^2\theta\right) \,,
\end{align}
where, for a given choice of reference axis for the azimuthal angle, the phase $\phi_0$ depends on the precise shape of the inclusion. Note that even though $\lambda$ is defined to be positive, $\bar{v}^\alpha(\br)/D \simeq \lambda J^{\alpha\beta}(\br) \hat{p}^\beta$ is formally left invariant under the joint transformation $\hat{\bp}\to - \hat{\bp}$ and $\lambda\to - \lambda$, therefore explaining why corrections to the $-2$ exponent in Eqs.~\eqref{eq:heur_res}-\eqref{eq:heur_res_2} appear only to second order in powers of $\lambda$.
\paragraph{Free obstacles:}
As discussed after Eq.~\eqref{eq:to_rescale}, the coupling to the fluid flow in Eq.~\eqref{eq:density_field} is irrelevant at large scales. The density field thus behaves as in a purely diffusive (dry) theory
\begin{equation}\label{eq:result_freely_moving}
    \delta\rho(\br) \simeq \frac{1}{4\pi D}\frac{r^\alpha}{r^3}\tilde{c}^\alpha \,,
\end{equation}
where $\tilde{c}^\alpha$ depends on the near-field details of the system and is generically non-zero for polar obstacles. The spatial decay exponent $-2$ is universal, and the non-universal vector $\tilde{c}^\alpha$ is contracted with a universal angular dependence. Note that for an obstacle with no polarity, even if fixed, we have by symmetry $\overline{\bF_{\rm{ext}}} = 0$ and so $\lambda = 0$. In this case, hydrodynamic effects are thus irrelevant on large scales, similarly to the case of freely-moving obstacles with arbitrary shape. Additionally, $\tilde{\bc}$ also vanishes by symmetry. We therefore expect density modulations to be governed by the next order term in the multipole expansion of the diffusion equation with a localized current at $\br = 0$, leading to $\delta\rho(\br) \sim r^{-3}$ at large distances. Also note that Eq.~\eqref{eq:result_freely_moving} strictly holds only if the orientation of the obstacle is constrained during motion. If its orientation rotates at a slow rate - either from fluctuations or from a ratchet effect - we expect the result in Eq.~\eqref{eq:result_freely_moving} to be screened beyond a lengthscale given by the typical distance run by diffusion during the persistence time of the orientation.

In the next sections, we derive the above results in a systematic manner starting from a microscopic model of spherical squirmers in the presence of a localized obstacle. 

\onecolumngrid

\twocolumngrid

\clearpage

\section{Microscopic model}\label{sec:micro}

We consider a fluid which obeys the Stokes equation
\begin{equation} \label{eq:stokes}
\eta \Delta \bv(\br) - \bnabla P(\br) = 0 \;, \;\;\; {\rm and} \;\;\; \bnabla \cdot \bv(\br) = 0 \,,
\end{equation}
where $\bv(\br)$ and $P(\br)$ are the flow and pressure fields at position $\br$. The fluid contains spherical squirmers of radius $a$, labeled by $i=1 \dots N$, with centers of mass at $\bx_i$. Each squirmer imposes, in a frame of reference moving with it, a velocity field $\bv_{s,i}(\br,\bu_i)$ on its surface. Here $\bu_i$ is a unit vector characterizing the orientation of the squirmer and we assume that $\bv_{s,i}(\br,\bu_i)$ has a polar asymmetry determined by $\bu_i$. We assume that the swimmers are dilute enough so that they interact only through hydrodynamics and that contact interactions between them can be neglected. The fluid also contains an obstacle that interacts with the swimmers both through hydrodynamics, by imposing a no-slip boundary condition on its surface, and directly through short-range external forces $\bF(\bx_i-\bx_0)$ and torques with respect to their center $\bGamma(\bx_i-\bx_0,\bu_i)$, with $\bx_0$ the center of mass of the obstacle. Denoting by $\dot{\bx}_0$ and $\bomega$ ($\dot{\bx}_i$ and $\bomega_i$) the translation and angular velocity of the obstacle (swimmer $i$), the above implies the boundary conditions on the surface of the obstacle, $\partial\Omega$,
\begin{equation} \label{eq:obsflow}
    \left. \bv(\br) \right|_{\partial\Omega} = \dot{\bx}_0 + \bomega \wedge \left(\br-\bx_0\right) \,,
\end{equation}
and on the surface of each swimmer, $\partial\Omega_i$,
\begin{equation}
    \left. \bv(\br) \right|_{\partial\Omega_i} = \dot{\bx}_i + \bomega_i \wedge \left(\br-\bx_i\right) + \bv_{s,i}(\br,\bu_i) \,.
\end{equation}
The translation and angular velocities $\dot{\bx}_i$ and $\bomega_i$ of the swimmers are such that the total force and torque exerted on each of them (by the fluid flow and the obstacle) vanish
\begin{equation}\label{eq:bc_force_swim}
    - \int_{\partial\Omega_i} \dd S \, n^{\mu} \sigma^{\mu\nu} + F^{\nu}(\bx_i-\bx_0) = 0 \,,
\end{equation}
and 
\begin{equation}
    - a \, \epsilon_{\alpha\mu\nu} \int_{\partial\Omega_i} \dd S \, n^{\mu} n^{\gamma} \sigma^{\gamma\nu} + \Gamma^\alpha(\bx_i-\bx_0, \bu_i) = 0 \,,
\end{equation}
where $\bn$ is an outward pointing normal vector to the surface of the swimmers, and $\sigma^{\mu\nu}(\br) = \eta \left( \partial_\mu v^\nu(\br) + \partial_\nu v^\mu(\br) \right) - P(\br)\delta^{\mu\nu}$ is the stress-tensor. We consider both the cases where the obstacle is held fixed externally, in which case $\dot{\bx}=0$ and $\bomega = 0$, and the case where it is free to move. For the latter, the force-free condition reads
\begin{equation} \label{eq:obsforce}
    - \int_{\partial\Omega} \dd S \, n^{\mu} \sigma^{\mu\nu} - \sum_i F_i^{\nu}(\bx_i-\bx_0) = 0 \,,
\end{equation}
and we assume that the motion is adiabatic so that the obstacle is much slower than the relaxation time of the squirmers' dynamics. In the remainder of this section, we compute the average far-field fluid flow generated by the swimmers suspension. We then use this average flow to build a mean-field model for the swimmers' dynamics, from which we recover Eq.~\eqref{eq:density_field}.

\subsection{The average fluid flow}
We start by computing the average fluid flow generated by the suspension. To do so, we use the boundary-integral representation of the Stokes equation, see Chapter 2 of \cite{pozrikidis1992boundary}, and express $\bv(\br)$ in terms of the velocity and stress-tensor at the boundary of the domain which is composed of the surfaces of the obstacle and of the swimmers. We obtain
\begin{widetext}
\begin{equation}\label{eq:fluid_flow_nonav}
\begin{split}
8 \pi \eta v^{\alpha}(\br) & =  \int_{\partial \Omega} \dd S \, n^{\rho} \sigma^{\rho\beta}\left(\br'\right)J^{\beta\alpha}\left(\br-\br'\right) - \eta \int_{\partial\Omega} \dd S \, v^\beta(\br') n^\gamma T^{\beta\gamma\alpha}\left(\br-\br'\right) \\ & + \sum_i \left[ \int_{\partial \Omega_i} \dd S \, n^{\rho} \sigma^{\rho\beta}\left(\br'\right)J^{\beta\alpha}\left(\br-\br'\right) - \eta \int_{\partial\Omega_i} \dd S \, v^\beta(\br') n^\gamma T^{\beta\gamma\alpha}\left(\br-\br'\right)\right]\,,
\end{split}
\end{equation}
\end{widetext}
where
\begin{equation}\label{eq:def_tensor_T}
T^{\alpha\beta\gamma}(\br) = - 6 \frac{r^\alpha r^\beta r^\gamma}{r^5} \,,
\end{equation}
generates the stress tensor corresponding to a Stokeslet solution and where $\br$' denotes the integration variable of the different surface integrals. While the velocity field $\bv(\br)$ is prescribed at the different surfaces over which the integrals are performed, the stress-tensor $\sigma^{\mu\nu}$ is not and, in principle, needs to be solved for. Equation \eqref{eq:fluid_flow_nonav} is thus implicit. It is nonetheless a useful starting point for determining the far-field flow. To proceed we use first the boundary conditions of the Stokes equation. From Eq. \eqref{eq:obsflow}, we note using Gauss's theorem that 
\begin{equation}\label{eq:vanish_surface}
\begin{split}
    & \int_{\partial\Omega} \dd S \, v^\beta(\br') n^\gamma T^{\beta\gamma\alpha}\left(\br-\br'\right) \\ & = \int_{\partial\Omega}\dd S \, n^\gamma T^{\beta\gamma\alpha}\left(\br-\br'\right) \left[\dot{x}^\beta_0 + \epsilon_{\beta\nu\delta}\,\omega^\nu(r'^\delta-x_0^\delta)\right] \\ & = - \int_\Omega \dd \br' \, \partial_\gamma T^{\beta\gamma\alpha}(\br-\br')\left[\dot{x}^\beta_0 + \epsilon_{\beta\nu\delta}\,\omega^\nu(r'^\delta-x_0^\delta)\right] \\ & + \int_\Omega \dd \br' \,\, \epsilon_{\beta\nu\delta}\,\omega^\nu \delta^{\delta\gamma}T^{\beta\gamma\alpha}\left(\br-\br'\right) \\ & = 0 \,,
\end{split}
\end{equation}
where we took advantage of the fact that $T^{\beta\gamma\alpha}(\br)$ is symmetric, see Eq.~\eqref{eq:def_tensor_T}, and that $\partial_\gamma T^{\beta\gamma\alpha}(\br - \br') = \delta^{\alpha\beta}\delta(\br-\br')$, which is a consequence of momentum conservation in the Stokes equation. Because the point $\br$ lies outside $\Omega$, this leads to the result of Eq.~\eqref{eq:vanish_surface}. Similar considerations also imply that 
\begin{align}
    & \int_{\partial\Omega_i}\dd S \, n^\gamma T^{\beta\gamma\alpha}\left(\br-\br'\right) v^\beta(\br') \nonumber \\ & = \int_{\partial\Omega_i}\dd S \, n^\gamma T^{\beta\gamma\alpha}\left(\br-\br'\right) v_{s,i}^\beta(\br', \bu_i)\,,
\end{align}
so that only the contribution from the surface velocity survives. Using these we obtain
\begin{equation}\label{eq:fluid_flow_mean_pos}
\begin{split}
8 \pi \eta v^{\alpha}(\br) & =  \int_{\partial \Omega} \dd S \, n^{\rho} \sigma^{\rho\beta}[\left\{\bx_i,\bu_i\right\}]\left(\br'\right)J^{\beta\alpha}\left(\br-\br'\right) \\ & + \sum_i \int_{\partial \Omega_i} \dd S \, n^{\rho} \sigma^{\rho\beta}[\left\{\bx_i,\bu_i\right\}]J^{\beta\alpha}\left(\br-\br'\right) \\ & - \sum_i \eta \int_{\partial\Omega_i}\dd S \, n^\mu T^{\mu\nu\alpha}(\br-\br') v_{s,i}^\nu(\br', \bu_i)\,.
\end{split}
\end{equation}
where the argument $\left\{\bx_i,\bu_i\right\}$ emphasizes that the stress-tensor $\sigma^{\rho\beta}\left(\br'\right)$ is a function of the positions and orientations of all the swimmers. 

We now evaluate the average flow $ \overline{v}^{\alpha}(\br)$, where the overline, as before, denotes an average over the many-body distribution $P[\left\{\bx_i,\bu_i\right\}]$ of the swimmers' positions and orientations. As noted previously, the motion of the obstacle is neglected. For simplicity, we thus consider $\bx_0 = 0$ in the following. For any point $\br'$ on the surface of the obstacle, we denote accordingly $\bar{\sigma}^{\rho\beta}_{\rm{obs}}(\br')$ the average stress-tensor at that point. Next, for any unit vector $\bn$, we introduce the average stress tensor on a swimmer's surface, at a location $a \bn$ with respect to its center
\begin{equation}
\bar{\sigma}^{\rho\beta}_{\rm{swim}}(\bx', \bx' + a \bn) \equiv \left\langle \sigma^{\rho\beta}[\left\{\bx_i,\bu_i\right\}]\left(\bx' + a \bn\right) \right\rangle_{\bx'}  \,,
\end{equation}
where we denote by $\left\langle \dots \right\rangle_{\bx'}$ a many-body average conditioned on the presence of a swimmer centered at $\bx'$, so that $\bx' + a \bn$ lies on the surface of one of the swimmers. Lastly, using the same notations, we introduce 
\begin{equation}
\bar{v}^{\nu}_{\rm{surf}}(\bx', \bx' + a \bn) \equiv \left\langle v_{s,j}^\nu\left(\bx' + a \bn, \bu\right) \right\rangle_{\bx'}\,, 
\end{equation}
the average surface velocity at $\bx' + a \bn$ on the surface of a swimmer centered at $\bx'$. Using these definitions and denoting by $\rho(\bx)=\langle \sum_i \delta(\bx-\bx_i) \rangle$ the mean density of swimmers, the average flow can thus be written as
\begin{widetext}
\begin{equation}\label{eq:fluid_flow1}
\begin{split}
& 8 \pi \eta \, \overline{v}^{\alpha}(\br) =  \int_{\partial \Omega} \dd S \, n^{\rho} \bar{\sigma}^{\rho\beta}_{\rm{obs}}(\br')J^{\beta\alpha}\left(\br-\br'\right) + \int \dd \bx' \rho(\bx') \int \dd \bn \, a^2 \, n^{\rho} \bar{\sigma}^{\rho\beta}_{\rm{swim}}(\bx', \bx' + a \bn)J^{\beta\alpha}\left(\br-\bx' - a \bn\right) \\ & -  \int \dd \bx' \rho(\bx') \int \dd \bn \, a^2 \, \eta \, n^\mu T^{\mu\nu\alpha}\left(\br-\bx' - a \bn\right) \bar{v}^{\nu}_{\rm{surf}}(\bx', \bx' + a \bn)\,.
\end{split}
\end{equation}
\end{widetext}
Equation \eqref{eq:fluid_flow1} can now be used for a multipole expansion. Since $T^{\mu\nu\alpha}(\br) \sim r^{-2}$ while $J^{\alpha\beta}(\br) \sim r^{-1}$, we obtain to leading order in the far field
\begin{align}\label{eq:fluid_flow_far_field}
& \overline{v}^{\alpha}(\br) \simeq \frac{1}{8\pi\eta} J^{\beta\alpha}\left(\br\right)\left[ \int_{\partial \Omega} \dd S \, n^{\rho} \bar{\sigma}^{\rho\beta}_{\rm{obs}}(\br') \right. \nonumber \\ & \left. + \int \dd \bx' \rho(\bx') \int \dd \bn \, a^2 \, n^{\rho} \bar{\sigma}^{\rho\beta}_{\rm{swim}}(\bx', \bx' + a \bn)\right] \,.
\end{align}
By definition $\int \dd \bn \, a^2 \, n^{\rho} \bar{\sigma}^{\rho\beta}_{\rm{swim}}(\bx', \bx' + a \bn)$ is minus the average force exerted by the fluid on a swimmer at position $\bx'$ and is therefore equal, using the force-balance condition in Eq.~\eqref{eq:bc_force_swim}, to the force exerted by the obstacle on that swimmer, that is $\int \dd \bn \, a^2 \, n^{\rho} \bar{\sigma}^{\rho\beta}_{\rm{swim}}(\bx', \bx' + a \bn) = F^\beta(\bx')$. We therefore get 
\begin{equation}\label{eq:fluid_flow_far_field2}
\begin{split}
 \overline{v}^{\alpha}(\br) & \simeq \frac{1}{8\pi\eta} J^{\beta\alpha}\left(\br\right)\left[ - \overline{ F^\beta_{\rm{fluid}\to\rm{obs}} } + \int \dd \bx' \rho(\bx') F^\beta(\bx')\right] \,,
\end{split}
\end{equation}
where $\overline{\bF_{\rm{fluid}\to\rm{obs}}} \equiv -\int_{\partial \Omega} \dd S \, n^{\rho} \bar{\sigma}^{\rho\beta}_{\rm{obs}}(\br')$ is the average force exerted by the fluid on the obstacle. The term between brackets thus reads, up to a minus sign, as the average for exerted by the fluid on the obstacle plus the average force exerted by the swimmers on the obstacle and is therefore equal to $\overline{\bF_{\rm{ext}}}$, the average external force exerted on the obstacle. This justifies Eq.~\eqref{eq:stokesflow}. As expected from the heuristic argument of Sec.~\ref{sec:heuristic}, a fixed obstacle embedded in a suspension of swimmers generates a far-field fluid flow that behaves as a Stokeslet. In addition, if the obstacle is (adiabatically) moving under force-free conditions, meaning that the total momentum of the system is conserved, the effective force monopole $\overline{\bF_{\rm{ext}}}$ vanishes. A higher order multipole expansion then shows that $\overline{ v^{\alpha}(\br)}$ behaves as the velocity field generated by a force dipole which decays as $r^{-2}$, see Eq.~\eqref{eq:dipoleflow}. The effective force dipole is given by
\begin{widetext}
\begin{equation}
\begin{split}
Q_{\rm{eff}}^{\beta\gamma} & =  \int_{\partial \Omega} \dd S \, n^{\rho} \bar{\sigma}^{\rho\beta}_{\rm{obs}}(\br')r'^\gamma + \int \dd \bx' \rho(\bx')x'^\gamma F^\beta(\bx') + \int \dd \bx' \rho(\bx') \int \dd \bn \, a^3 \, n^{\rho}n^{\gamma} \bar{\sigma}^{\rho\beta}_{\rm{swim}}(\bx', \bx' + a \bn) \\ & + \int \dd \bx' \rho(\bx') \int \dd \bn \, a^2 \, \eta \, \left[n^\gamma  \bar{v}^{\beta}_{\rm{surf}}(\bx', \bx' + a \bn) +  n^\beta  \bar{v}^{\gamma}_{\rm{surf}}(\bx', \bx' + a \bn)\right] \,.
\end{split}
\end{equation}
\end{widetext}

\subsection{Mean-field approximation}\label{sec:mean_field}
With the expression for the mean flow at hand, we can now turn to derive the drift-diffusion equation Eq.~\eqref{eq:density_field}. We use a mean-field approximation where we consider the motion of a single swimmer in a steady inhomogeneous background flow identified with the average flow $\bar{\bv}(\bx)$ derived above. For that swimmer, the equations of motion read
\begin{equation}\label{eq:translation}
    \dot{\bx} = \mu \bF(\bx) + v_0 \bu + \bar{\bv}(\bx) 
\end{equation}
together with
\begin{equation}\label{eq:rotation}
    \dot{\bu} = \left( \mu_r \bGamma(\bx,\bu) + \frac{1}{2} \bnabla \wedge \bar{\bv}(\bx) \right) \wedge \bu + \text{noise} \,,
\end{equation}
where the noise is taken for simplicity to be of the run-and-tumble type~\footnote{The same derivation can be repeated with rotational diffusion with the conclusions unchanged.}. Here $\mu = 1/(6\pi\eta a)$ is the mobility of a sphere of radius $a$ and $\mu_r = 1/(8\pi\eta a^3)$ is the corresponding rotational mobility. Also, $v_0$ is the self-propulsion speed of an isolated swimmer which is given by
\begin{equation}
    v_0 = - \frac{1}{4 \pi a^2}\int \dd S \, \bv_s(\br,\bu) \cdot \bu \,.
\end{equation}
Henceforth, to ease the notations, we use $\bar{\bomega}(\bx) \equiv (1/2) \bnabla \wedge \bar{\bv}(\bx)$. These equations have been derived in \cite{stone1996propulsion} in the absence of an external force $\bF=0$ and torque $\bGamma = 0$ and in the absence of a background flow $\bar{\bv}(\bx) = 0$. The results of \cite{stone1996propulsion} generalize to Eqs.~\eqref{eq:translation}-\eqref{eq:rotation}, as we show in Appendix~\ref{app:single_swimmer}, for swimmers much smaller than the scale of variation of $\bar{\bv}(\bx)$.

Our interest is in the steady-state density profile generated by the dynamics in Eqs.~\eqref{eq:translation}-\eqref{eq:rotation}. Let $\psi(\bx,\bu)$ be the steady-state distribution. It is a solution of
\begin{equation} \label{eq:FP1body}
\begin{split}
    0 = & - \bnabla_{\bx} \cdot \left( \left[\mu \bF(\bx) + v_0 \bu + \bar{\bv}(\bx)\right] \psi(\bx,\bu) \right) \nonumber \\ &  + \frac{1}{\tau} \sum_i \left( \int \dd \bu' \psi(\bx,\bu') - \psi(\bx,\bu)\right) \nonumber \\ & - \bnabla_{\bu} \cdot \left( \left[\left( \mu_r \bGamma(\bx,\bu) + \bar{\bomega}(\bx) \right) \wedge \bu\right] \psi(\bx,\bu) \right) \,.
\end{split}
\end{equation}
We introduce the density $\rho(\bx)=\int \dd \bu \, \psi(\bx,\bu)$, polarity $m^\mu(\bx)=\int \dd \bu \, u^\mu \, \psi(\bx,\bu)$ and nematic tensor $Q^{\alpha\beta}(\bx)=\int \dd \bu \left(u^\alpha u^\beta - \frac{\delta^{\alpha\beta}}{3}\right)\psi(\bx,\bu)$.
Upon integrating Eq.~\eqref{eq:FP1body} over $\bu$, we get
\begin{equation}\label{eq:den}
    - \partial_\alpha \left[ \mu F^{\alpha}(\bx) \rho(\bx) + v_0 m^{\alpha}(\bx) + \bar{v}^{\alpha}(\bx)\rho(\bx) \right] = 0 \,.
\end{equation}
Multiplying Eq.~\eqref{eq:FP1body} by $u^\beta$ and integrating it again over $\bu$ yields
\begin{align}\label{eq:polarization}
    & \frac{m^\beta}{\tau} = - \frac{v_0}{3}\partial_\beta \rho - \partial_{\alpha}\left[\mu F^{\alpha}m^\beta + v_0 Q^{\alpha\beta} + \bar{v}^{\alpha}(\bx)m^\beta(\bx) \right] \nonumber \\ & + \epsilon_{\beta\mu\nu}\left[\mu_r \int \dd \bu \, u^\nu \Gamma^\mu(\bx,\bu) \psi(\bx,\bu) + \bar{\omega}^{\mu}(\bx) m^{\nu}(\bx) \right]\,,
\end{align}
which can be used in Eq. \eqref{eq:den} to give
\begin{widetext}
\begin{equation}
\begin{split}
    &\frac{v_0^2 \tau}{3}\partial_\alpha \partial^\alpha \rho(\bx) - \partial_\alpha \left[\bar{v}^\alpha(\bx) \rho(\bx) \right] = \partial_\alpha \left\{\mu F^{\alpha}(\bx) \rho(\bx) + v_0\tau \epsilon_{\alpha\mu\nu}\left[\mu_r \int \dd \bu \, u^\nu \Gamma^\mu(\bx,\bu) \psi(\bx,\bu) + \bar{\omega}^{\mu}(\bx) m^{\nu}(\bx) \right]\right\} \\ & - v_0\tau  \partial^\alpha \partial^\beta \left[\mu F^{\alpha}m^\beta + v_0 Q^{\alpha\beta} + \bar{v}^{\alpha}(\bx)m^\beta(\bx)\right] \,.
\end{split}
\end{equation}
\end{widetext}
Therefore we find that the equation satisfied by the density field can be written as a drift-diffusion equation with sources as in Eq.~\eqref{eq:density_field}, where $C^\alpha(\bx) = C_1^\alpha(\bx) + C_2^\alpha(\bx)$ with
\begin{align}
   C_1^\alpha(\bx) = & - \mu F^{\alpha}(\bx) \rho(\bx) + v_0\tau \mu  \partial^\beta \left[ F^{\alpha}(\bx)m^\beta(\bx) \right] \nonumber \\ & - v_0\tau \mu_r \epsilon_{\alpha\mu\nu} \int \dd \bu \, u^\nu \Gamma^\mu(\bx,\bu) \psi(\bx,\bu) \,,
\end{align}
and 
\begin{align}
   C_2^\alpha(\bx) = & - v_0\tau \epsilon_{\alpha\mu\nu} \bar{\omega}^{\mu}(\bx) m^{\nu}(\bx)\nonumber \\ & + v_0\tau   \partial^\beta \left[ v_0 Q^{\alpha\beta} + \bar{v}^{\alpha}(\bx)m^\beta(\bx)\right] \,.
\end{align}
It is clear that the integral of $C_1^\alpha(\bx)$ is finite since the force and torque fields $\bF(\bx)$ and $\bGamma(\bu,\bx)$ are short-ranged. To bridge the gap with Eq.~\eqref{eq:density_field} and the discussion in Sec.~\ref{sec:heuristic}, we now argue that 
\begin{equation}\label{eq:defc}
c_2^\alpha = \int \dd \bx \, C_2^{\alpha}(\bx)  
\end{equation}
is also finite. Since we cannot solve the whole hierarchy of angular moments, we proceed by self-consistency assuming that $c_2^\alpha$ exists. As we have discussed in Sec.~\ref{sec:heuristic} and is shown in the following section, the density field decays faster than $x^{-1}$. The polarity $m^\alpha(\bx)$ then decays faster than $x^{-2}$, since it is proportional to density gradients, see Eq.~\eqref{eq:polarization}. Accordingly, we expect that $Q^{\alpha\beta}(\bx)$ decays faster than $O(x^{-3})$. In fact, successive moments of the orientation decay faster and faster, which can be shown in any truncation of the hierarchy of angular moments. Therefore, we expect that $C_2^\alpha(\bx)$ decays faster than $x^{-4}$ and is indeed integrable, thereby closing the self-consistency argument.

\section{Far-field decay of the density field}\label{sec:asympt}

In this section, we derive the far-field density decay when the obstacle is held fixed. To do so we use a similarity solution, close to what is done, for example, for the Barenblatt equation, see Chapter 10  of \cite{goldenfeld2018lectures} and Chapter 3 of \cite{barenblatt1996scaling}. Even though the Barenblatt equation features a time dependence that ours does not, in both cases, the large-scale behavior of the partial differential equation under study is mapped, by choosing a suitable ansatz, to an ordinary differential equation from which the anomalous exponent is obtained by solving a non-linear eigenvalue problem. For completeness, the same results are derived using a renormalization group procedure in Appendix~\ref{app:RG}. In the far field, we look for a solution of 
\begin{equation}\label{eq:density_far}
    D \Delta \delta\rho - \bnabla \cdot \left[\bar{\bv}(\br) \delta\rho \right] = - \bc\cdot\bnabla \delta(\br) \,,
\end{equation}
where the convective flow, derived in Eq.~\eqref{eq:fluid_flow1}, follows the scale-free form given in Eq.~\eqref{eq:stokesflow} at large distances. We work with spherical coordinates with polar angle $\theta$ such that $\cos\theta = \hat{\bp} \cdot \hat{\br}$, where $\hat{\bp}$, defined in Eq.~\eqref{eq:pmono}, points along the force monopole, and with an azimutal angle $\phi$. Dimensional analysis then shows that 
\begin{equation}\label{eq:scaling_form}
    \delta\rho(\br) = \frac{1}{r^2}\frac{|\bc|}{D}\mathcal{F}\left(\frac{\ell}{r}, \theta, \phi\right) \,,
\end{equation}
where $\ell$ is a microscopic length scale emerging from the near-field behavior of the velocity field. We first decompose $\mathcal{F}$ into Fourier modes
\begin{equation} \label{eq:denseries}
    \mathcal{F}\left(\frac{\ell}{r}, \theta, \phi\right) = \sum_{m=-\infty}^{+\infty}  \ee^{i m \phi}\,f_m\left(\frac{\ell}{r}, \theta\right)\,.
\end{equation}
In the far-field, with $r$ much larger than any microscopic length scale, we write each Fourier mode as a product $f_m(\ell/r, \theta) \propto g_m(\theta)r^{-\epsilon_m}$ and we find using Eq.~\eqref{eq:to_rescale} that the angular functions satisfy
\begin{widetext}
\begin{equation}\label{eq:ang_sol}
    \frac{1}{\sin\theta}\partial_\theta\left(\sin\theta \, \partial_\theta g_m \right) + \lambda \sin\theta \partial_\theta g_m + g_m \left[(2+\epsilon_m)(1+\epsilon_m) + 2 \lambda(2+\epsilon_m)\cos\theta - \frac{m^2}{\sin^2\theta}\right] = 0 \,,
\end{equation}
\end{widetext}
where $\lambda$ is defined in Eq.~\eqref{eq:lambda}. The exponent $\epsilon_m$ is then fixed by requiring that Eq.~\eqref{eq:ang_sol} has a well-behaved solution at the boundaries of the interval $\cos\theta = \pm 1 $. For a freely-moving obstacle, meaning when $\lambda = 0$, or equivalently in the absence of hydrodynamic interactions, the set of possible exponents $\epsilon_m$ are integers such that $\epsilon_m \geq |m| - 1$. Since the source term in Eq.~\eqref{eq:density_far} is a derivative of a delta function, the far-field decay of the density field is dominated by the modes $m=0$ and $m=\pm 1$, with exponents $\epsilon_{0,\pm 1} = 0$, meaning $\delta\rho(\br) \sim r^{-2}$. The solution $\epsilon_0=-1$ is indeed ignored as it corresponds to a delta function source. This reproduces the well-known Eq.~\eqref{eq:result_freely_moving} for the solution of the Laplace equation in the presence of a localized current.

When $\lambda > 0$ and small, the far-field decay of the density field is also dominated by the modes $m=0,\pm1$. Indeed, as will be clear in the following, higher modes $|m| \geq 2$ correspond to decay exponents close to $|m| - 1 \geq 1$ when $\lambda$ is small and therefore contribute only as subleading corrections in the far-field compared to the modes $m = 0, \pm 1$. 

To characterize these modes, it is naively tempting to postulate $\epsilon_{0,\pm 1}=0$ and solve for $g_m(\theta)$ using a perturbation theory in $\lambda$. However, solutions of this form inevitably diverge at one of the endpoints $\cos\theta = \pm 1$, to order $O(\lambda^2)$, as we show in Appendix~\ref{app:singular}. This signals the presence of an anomalous exponent $\epsilon_m \neq 0$.

We now evaluate the exponents $\epsilon_m$ and the angular functions $g_m(\theta)$ perturbatively in $\lambda$ using $\epsilon_m = \lambda \epsilon_m^{(1)} + \lambda^2 \epsilon_m^{(2)}  +O(\lambda^3)$ and $g_m(\theta) = g_m^{(0)}(\theta) + \lambda g_m^{(1)}(\theta) + \lambda^2 g_m^{(2)}(\theta)+O(\lambda^3)$. Requiring that $g_m(\theta)$ remains finite to second order in $\lambda$ at $\cos\theta=\pm 1$ yields the anomalous exponents
\begin{eqnarray} \label{eq:exponpert}
    \epsilon_0 &=& \frac{1}{3}\lambda^2 + O\left(\lambda^4\right) \,, \\
    \epsilon_{\pm 1} &=& - \frac{1}{12}\lambda^2 + O(\lambda^4) \nonumber \,,
\end{eqnarray}
and the angular function to order $O(\lambda ^2)$,
\begin{eqnarray} \label{eq:angpert}
      g_0(\theta) &\propto& \cos\theta - \frac{\lambda}{4}  \left(3-5 \cos^2\theta\right) + \frac{3 \lambda ^2 \cos^3\theta}{4} \,, \nonumber \\
      g_{\pm 1}(\theta) &\propto& \sin\theta\left(1 + \frac{5}{4}\lambda \cos\theta + \frac{3}{4}\lambda^2 \cos^2\theta\right) \,. 
\end{eqnarray}
The above equations can then be used to obtain the results presented in Sec. \ref{sec:heuristic}. For a generic polar obstacle, the far-field density is governed by the slowest $m=\pm 1$ modes and we identify $\epsilon_\perp \equiv \epsilon_{\pm1}$. We thus recover Eqs.~\eqref{eq:heur_res_2} and \eqref{eq:heur_res_2bis}, where in Eq. (\ref{eq:heur_res_2bis}) the dependence on the azimuthal angle from Eq.~\eqref{eq:denseries} is included. In contrast, if the obstacle possesses an axis of symmetry, whose direction $\hat{\bp}$ must be pointing along \footnote{If the obstacle has an axis of rotational symmetry, and the rest of the system is completely isotropic (which assumes that the bulk suspension does not exhibit orientational order), then any non-vanishing vector built from the steady-state distributions of stresses and positions of the active particles must be along this axis.}, the modes $m=\pm 1$ must vanish, and the far-field decay is thus governed by the $m=0$ mode. This holds whether the rotational symmetry is continuous or discrete. Hence, we identify $\epsilon_\parallel\equiv\epsilon_{0}$ and get Eqs.~\eqref{eq:heur_res} and \eqref{eq:heur_res_bis}. It is in principle straightforward to extend this procedure to arbitrary order in $\lambda$.

\section{Interactions between bodies} \label{sec:interactions}
Since an inclusion generates a long-range density modulation and a long-range fluid flow in the system, it affects the neighborhood of other inclusions. This leads to long-range interactions, mediated by the swimmers and the viscous fluid, that we explore in this section. Such long-range mediated interactions are well-known between particles, passive or active, embedded in a viscous fluid \cite{felderhof1977hydrodynamic, lauga2009hydrodynamics} and have been recently calculated for passive inclusions in ``dry'' active systems \cite{baek2018generic}. In the case we consider here, both the hydrodynamic field and the active particles mediate the interactions. 

In this section, we derive the long-range mediated interactions that emerge between two inclusions immersed in a three-dimensional suspension of self-propelling particles, in two simple cases. First, we describe the dynamics (within an adiabatic approximation) of two inclusions that are pinned at one point but free to rotate around this point. Second, we discuss the effective interactions between two freely moving inclusions. We assume that the inclusions are polar and, for simplicity, with an axis of symmetry. The extension to other cases is straightforward even if tedious.

\subsection{Two Fixed Polar Obstacles}
We consider two fixed inclusions, at position $\br_1$ and $\br_2$, and denoted in the following by 1 and 2. Asymptotically, when the distance $|\br_1-\br_2|$ goes to infinity, each inclusion has to be held in place by an average force, denoted $\bar{\bF}_1$ for inclusion 1 and $\bar{\bF}_2$ for inclusion 2, in order to maintain their position fixed. Note that due to the axisymmetry of the obstacles, there is no need to exert an average torque in order to prevent them from rotating.

\begin{figure}[h]
\centering
\includegraphics[width=\linewidth]{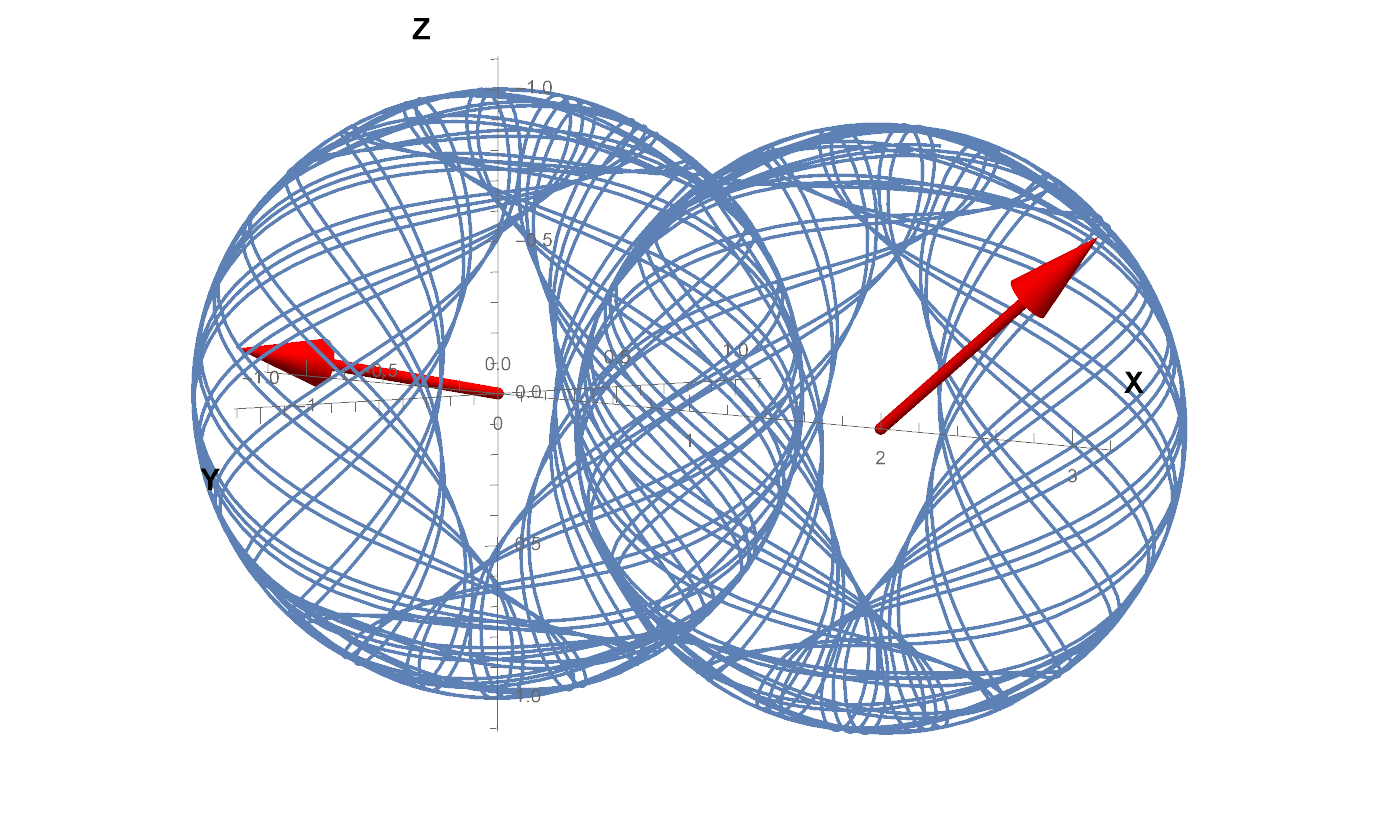}
\caption{Examples of complex trajectories induced by the interactions between an aligning ($\gamma_1 >0$) and an anti-aligning ($\gamma_2 < 0$) freely-rotating polar object embedded in a suspension of microswimmers. The center of the first obstacle is located at the origin and that of obstacle 2 is on the $x$-axis. The instantaneous position of the two directors $\hat{\bp}_1$ and $\hat{\bp}_2$ at some time $t>0$ is depicted by two red arrows while the solid blue lines represent the past trajectories starting from a random initial condition at $t=0$. For such generic initial conditions, the trajectory of each director seem to densely cover a portion of the sphere at large times. Here $\gamma_1 = - \gamma_2$.}
\label{fig:interactions}
\end{figure}

We now consider a case where these two obstacles are pinned at points $\br_1$ and $\br_2$, but each free to rotate around that pinning point. We assume that the pinning points lie on the axis of symmetry of the corresponding inclusion. When $|\br_1 - \br_2|$ is large but finite, the presence of obstacle 1 induces a far-field fluid flow around obstacle 2, which influences its orientation. We treat the dynamics within the adiabatic approximation so that at each time the two inclusions behave as fixed force monopoles, and we use the conventions $\hat{\bp}_2(t) = \bar{\bF}_2(t)/|\bar{\bF}_2|$, $\hat{\bp}_1(t) = \bar{\bF}_1(t)/|\bar{\bF}_1|$ and $\hat{\br}_{21}=\left(\br_2-\br_1\right)/|\br_2-\br_1|$. Neglecting fluctuations, the dynamics of the orientation of the first inclusion reads
\begin{equation}
    \frac{\dd \hat{\bp}_1}{\dd t} = \overline{\bomega}_1\left(\hat{\bp}_1,\hat{\bp}_2,\br_1,\br_2\right) \wedge \hat{\bp}_1 \,,
\end{equation}
where $\overline{\bomega}_1\left(\hat{\bp}_1,\hat{\bp}_2,\br_1,\br_2\right) $ is the average angular velocity of obstacle 1 at orientation $\hat{\bp}_1$ in the presence of (the far-away) obstacle 2 with fixed orientation $\hat{\bp}_2$. The impact resulting from variations in swimmer density modulations (scaling as $\sim |\br_2-\br_1|^{-2 + \epsilon}$) is minimal when compared to the fluid flow (scaling as $\sim |\br_2-\br_1|^{-1}$), at least perturbatively in $\lambda$. Therefore, to leading order in the distance $|\br_1 - \br_2|$, the angular velocity can be expressed as a linear response to the Stokeslet flow $\bv_2\left(\hat{\bp}_2, \br_1,\br_2\right)$ generated by obstacle 2 at point $\br_1$ in the absence of obstacle 1, as in \cite{chajwa2020waves}, 
\begin{equation}
  \overline{\omega}^\mu_1\left(\hat{\bp}_1,\hat{\bp}_2,\br_1,\br_2\right)  = M_1^{\mu \nu}\left(\hat{\bp}_1\right)v^\nu_2\left(\hat{\bp}_2, \br_1,\br_2\right) \,.
\end{equation}
 Here $M_1^{\mu \nu}\left(\hat{\bp}_1\right)$ is the linear-response tensor of the average angular velocity of obstacle 1 to a uniform background flow. Note that the pinning of obstacle 1 breaks Galilean invariance, therefore coupling the dynamics of $\hat{\bp}_1(t)$ to the fluid flow $\bv_2\left(\hat{\bp}_2, \br_1,\br_2\right)$ itself and not only to its gradients (other instances in which Galilean invariance is explicitly broken in active suspensions, therefore leading to possible alignment with the local suspension velocity, include confined suspensions and suspensions on substrates \cite{maitraPRL2020,brottoPRL2013,kumar2014flocking}). By symmetry, the linear-response tensor must be antisymmetric in the indices $(\mu,\nu)$ and invariant under rotations around $\hat{\bp}_1$. This yields
\begin{equation}
   M_1^{\mu \nu}\left(\hat{\bp}_1\right) = -\gamma_1 \epsilon_{\mu\nu\alpha} \hat{p}_1^\alpha \,,
\end{equation}
with $\gamma_1$ an object-dependent coefficient that depends on the near-field properties of the active suspension in the vicinity of obstacle 1. Note that $\gamma_1 > 0$ implies that in a steady uniform background flow, $\hat{\bp}_1$ aligns with the flow, while it anti-aligns with it if $\gamma_1 < 0$. Therefore, one has
\begin{equation}\label{eq:dyn_p1}
    \frac{\dd \hat{\bp}_1}{\dd t} = \frac{\gamma_1 |\bar{\bF}_2|}{8 \pi \eta |\br_1 - \br_2|} \, \hat{\bp}_1 \wedge \Big[\left(\hat{\bp}_2 + \left(\hat{\bp}_2\cdot \hat{\br}_{12}\right)\hat{\br}_{12}\right) \wedge \hat{\bp}_1\Big] \,.
\end{equation}
Accordingly, the dynamics of $\hat{\bp}_2(t)$ follows from 
\begin{equation}\label{eq:dyn_p2}
    \frac{\dd \hat{\bp}_2}{\dd t} = \frac{\gamma_2 |\bar{\bF}_1|}{8 \pi \eta |\br_1 - \br_2|} \, \hat{\bp}_2 \wedge \Big[\left(\hat{\bp}_1 + \left(\hat{\bp}_1\cdot \hat{\br}_{12}\right)\hat{\br}_{12}\right) \wedge \hat{\bp}_2\Big] \,.
\end{equation}
The lack of reciprocity in the interactions between the two inclusions visible in Eqs.~(\ref{eq:dyn_p1}, \ref{eq:dyn_p2}) is a trademark of interactions mediated by active baths \cite{baek2018generic,saha2019pairing, granek2020bodies}. When $\gamma_1 > 0$ and $\gamma_2 >0$, the effective interactions drive alignment between the two directors in the direction separating the two inclusions, meaning $\bp_1 = \bp_2 = \pm \hat{\br}_{12}$ in the steady-state. Furthermore, when both $\gamma_1 < 0$ and $\gamma_2 < 0$, the effective interactions lead to anti-alignment between the two directors in the direction separating the two inclusions, meaning $\bp_1 = -\bp_2 = \pm \hat{\br}_{12}$. None of these equilibrium points is stable when $\gamma_1 \gamma_2 < 0$. In fact, numerical solutions of the joint dynamics Eqs.~(\ref{eq:dyn_p1}, \ref{eq:dyn_p2}) show that interactions between two such freely-rotating bodies generically lead to complex trajectories of $\hat{\bp}_1$ and $\hat{\bp}_2$, see Fig.~\ref{fig:interactions}. The dynamics are rich depending on the initial conditions and their study, including the influence of noise on the dynamics Eqs.~(\ref{eq:dyn_p1}, \ref{eq:dyn_p2}) or in the presence of more than two bodies, is left for future work.

Broadly speaking, the above phenomenology was already identified in the dynamics of pinned inclusions in suspensions of dry active particles \cite{baek2018generic,granek2020bodies}, albeit with slightly different dynamics. We stress however that momentum conservation leads to much longer-ranged effective interactions. In fact, the effective interactions in Eqs.~(\ref{eq:dyn_p1}, \ref{eq:dyn_p2}) decay as $O(|\br_1-\br_2|^{-1})$, whereas they were shown to decay as $O(|\br_1-\br_2|^{-3})$ in three-dimensional dry systems \cite{baek2018generic}. This difference could have striking consequences on the behavior of ensembles of pinned embedded inclusions.

\vspace{0.5cm}
\vspace{0.5cm}

\subsection{Freely-moving bodies}
Next, consider the case of two freely-moving obstacles. Let $\bu_1$ ($\bu_2$) denote the average velocity of obstacle 1 (obstacle 2) when in isolation. Then, the far-field density decay around obstacle 2 follows from Eq.~\eqref{eq:result_freely_moving} and reads
\begin{equation}
    \rho(\br) = \rho_0 + \delta\rho_2(\br) \,\,\, \rm{with} \,\,\, \delta\rho_2(\br) \simeq \frac{1}{4\pi D}\frac{(\br-\br_2)\cdot\tilde{\bc}_2}{|\br-\br_2|^3} \,.
\end{equation}
The same result holds around obstacle 1 upon replacing $\tilde{\bc}_2$ by $\tilde{\bc}_1$ and $\br_2$ by $\br_1$. For what follows we introduce $v_1(\rho_0)$ and $v_2(\rho_0)$ the average speed of obstacles 1 and 2 respectively, which are scalar functions of the bulk density $\rho_0$.

There are two sources for the interaction between the inclusions. First, there is a contribution from the fluid flow created by one inclusion in the vicinity of the other. The other one comes from the change in swimmers' density in the vicinity of one inclusion due to the presence of the other. Both contributions scale in the same manner with the distance between the inclusions.

We denote the changes in the average velocity of each obstacle by $\bu_1 + \delta \bu_1$ and $\bu_2 + \delta \bu_2$ for obstacles 1 and 2 respectively. To leading order in the far field, $\delta \bu_1$ is given by the sum of the two contributions discussed above. First, due to the presence of object 2, the apparent bulk density of swimmers around obstacle 1 is perturbed, going from $\rho_0$ to $\rho_0 + \delta\rho_2(\br_1)$. This scalar perturbation modifies the speed of obstacle 1, but not the propulsion direction. The second contribution emerges from the coupling to the fluid flow generated by object 2 which behaves as the one generated by a force dipole $Q_2^{\alpha\beta}$ at position $\br_2$. These two contributions scale as $|\br_1-\br_2|^{-2}$ and yield
\begin{equation}
        \delta u_1^\alpha = \frac{1}{8\pi\eta}\partial_\gamma J^{\alpha\beta}(\br_1 - \br_2) Q_2^{\gamma\beta} + \hat{u}_1^\alpha v_1'(\rho_0)\delta\rho_2(\br_1) \,.
\end{equation}
Because obstacle 2 is polar with an axis of symmetry, we have $\tilde{\bc}_2 = \chi_2 \hat{\bu}_2$ with $\chi_2$ a parameter which depends on near-field properties of the suspension close to obstacle 2. Furthermore, we have
\begin{equation}
        Q_2^{\gamma\beta} = \kappa_2 \left(\hat{\bu}_2^\gamma \hat{\bu}_2^\beta - \frac{\delta^{\gamma\beta}}{3}\right) \,,
\end{equation}
with $\kappa_2$ also depending on the near-field properties of the suspension close to obstacle 2.
Hence, to leading order, the effective interactions between the two bodies take the form 
\begin{align}
        \delta u_1^\alpha = & -\frac{\kappa_2 \, r_{21}^\alpha}{8\pi\eta |\br_2 - \br_1|^2}\left(1-3\left(\hat{\br}_{12} \cdot \hat{\bu}_2\right)^2\right) \nonumber \\ & - \chi_2 \hat{u}_1^\alpha\frac{v_1'(\rho_0)}{4 \pi D}\frac{\hat{\br}_{21}\cdot \hat{\bu}_2}{|\br_1-\br_2|^2}\,,
\end{align}
and correspondingly for the shift $\delta\bu_2$ in the velocity of object 2. The first term is a swimmer-swimmer interaction, showing that passive bodies embedded in an active suspension partly behave as swimming particles themselves. The second term however does not correspond to a swimmer-swimmer interaction but is akin to the far-field interactions emerging between two passive bodies embedded in a medium of ``dry'' self-propelled particles \cite{baek2018generic}.

\section{Conclusion}
In this paper, we studied the long-range effect of a localized obstacle on a three-dimensional suspension of active swimmers. First, we showed that hydrodynamic interactions can lead to striking deviations from earlier results obtained in the dry case when the obstacle is held fixed by an external force so that there is a net average flux of momentum injected into the system. In that case, the far-field density modulations of the swimmers decay with an exponent that depends continuously on the relative amplitude of hydrodynamic and diffusive contributions. The exponent also depends on the internal symmetry of the obstacle: a polar obstacle with an axis of symmetry induces density modulations that decay faster than in the absence of hydrodynamic interactions while an obstacle with no axis of symmetry induces modulations that decay slower than in the dry case. In both cases, we have a perturbative prediction for the exponent in terms of the independently measurable quantities $|\overline{\bF}_{\rm{ext}}|$, $\eta$ and $D$. In particular, $|\overline{\bF}_{\rm{ext}}|$ can be read off from the leading far-field decay of the hydrodynamic velocity. The case of a freely-moving inclusion is closer to earlier studies on the dry problem. There, hydrodynamic interactions are irrelevant far away from the obstacle, and the $-2$ exponent is recovered \cite{baek2018generic}. As argued in Sec.~\ref{sec:heuristic}, these predictions emerge from a competition between diffusive effects and convective transport due to the local injection of momentum in the vicinity of the obstacle. We believe this scenario is generic enough for our results to robustly extend beyond the presently studied case of spherical squirmers and be appraised in experiments on synthetic or biological microswimmers. We stress that our predictions rely on the three-dimensional nature of the surrounding fluid flow. In fact, in the vicinity of a container's wall,  acting as a momentum sink, the flow field around a localized momentum source decays faster (as $\sim 1/r^2$) when compared to three-dimensional bulk fluids. In such a case, following the dimensional analysis of Eq.~\eqref{eq:to_rescale},
we therefore expect hydrodynamic interactions to be irrelevant far away from a localized obstacle, even if it is held fixed. Note however that we expect effects similar to the ones described here if the motion of the microswimmers is limited near the interface between two immiscible viscous fluids or inside a two-dimensional fluid layer in a three-dimensional viscous fluid.

In addition, we have also described the effective long-range interactions, mediated by the active suspension, between two far-away localized objects. If freely moving, the effective interactions between the two objects lead to a modification of their average propulsion velocity. This modification decays as the distance between the two objects squared and can be expressed as the sum of two contributions. The first one is akin to the hydrodynamic interactions existing between two force dipoles. The second contribution has the same form as the effective interactions mediated by a bath of ``dry'' self-propelled particles \cite{baek2018generic}. When their center of mass is held fixed, effective torques emerge, that decay as the inverse of the distance between the two obstacles. Depending on the details these can either lead to alignment, anti-alignment, or complex trajectories.

We believe this study opens the way for a quantitative description of many phenomena, including the effect of disorder on suspensions of microswimmers~\cite{dor2019ramifications,ro2021disorder,dor2022disordered,granek2023inclusions}, and the interactions of inclusions with confining walls~\cite{dor2022passive}.

\begin{acknowledgments}
We thank Ram Adar, Mehran Kardar, and Ananyo Maitra for discussions and comments on the manuscript. TADP and YK acknowledge financial support from ISF (2038/21) and NSF/BSF (2022605), and SR from a JC Bose Fellowship of the SERB, India and an ICTS Simons Visiting Professorship of the \href{https://www.icts.res.in/}{International Centre for Theoretical Sciences}. TADP thanks the Laboratoire de Physique Th\'eorique et Hautes Energies at Sorbonne Universit\'e for hospitality. SR thanks the Isaac Newton Institute for Mathematical Sciences, Cambridge, for support and hospitality during the programme \href{https://www.newton.ac.uk/event/adi/}{Anti-diffusive dynamics: from sub-cellular to astrophysical scales} where work on this paper was undertaken, supported by EPSRC grant no EP/R014604/1. This research began in interactions during the \href{https://www.icts.res.in/}{ICTS-TIFR} program \href{https://www.icts.res.in/discussion-meeting/SPCS2022}{Statistical Physics of Complex Systems (SPCS2022)}, for whose support and hospitality we are grateful.
\end{acknowledgments} 

\clearpage

\onecolumngrid
\appendix

\section{Dynamics of an isolated swimmer}\label{app:single_swimmer}
The dynamics of an isolated squirmer, a spherical particle that self-propels in a viscous fluid by imposing a non-zero surface flow in its frame of reference, has been derived in \cite{stone1996propulsion}. In this appendix, we extend their derivation to the case where an external force and torque are imposed on the squirmer. Because of the linearity of the Stokes equation, the resulting velocity is the sum of the self-propulsion of the isolated squirmer and of the translation velocity of a passive sphere of the same size driven by the external force. 

The squirmer motion is a combination of translation with velocity $\dot{\bx}$ and solid rotation with angular velocity $\bomega$. The equation governing the fluid flow reads
\begin{equation}\label{eq:stokes_app1}
\eta \Delta \bv - \bnabla P = 0
\end{equation}
together with 
\begin{equation}\label{eq:div_app1}
    \bnabla \cdot \bv = 0 \,,
\end{equation}
and the boundary conditions
\begin{equation}\label{eq:bc_app1}
    \left. \bv \right|_{\partial\Omega}(\br) = \dot{\bx} + \bv_{s}(\br,\bu) + a \, \bomega \wedge \bn \,\,\, \text{and} \,\,\, \left. \bv \right|_{\infty} = 0 \,,
\end{equation}
with $\bv_{s}(\br,\bu)$ the local surface velocity imposed by the swimmer in its frame of reference and $\bn$ is the local outward-pointing normal to the squirmer's surface $\partial\Omega$. We recall that $\bv_{s}(\br,\bu)$ has a polarity, that is, a vectorial asymmetry,
determined by $\bu$. The translation velocity $\dot{\bx}$ is fixed by the force-balance condition
\begin{equation}
\int_{\partial\Omega} \dd S n^{\beta} \sigma^{\alpha\beta}\left(\br'\right) = F^\alpha \,,
\end{equation}
and the angular velocity $\bomega$ is fixed by the torque-balance condition
\begin{equation}
a \, \epsilon_{\rho\alpha\beta} \int_{\partial\Omega} \dd S n^{\alpha} n^{\mu} \sigma^{\mu\beta} = \Gamma^\rho \,.
\end{equation}
In order to obtain $\dot{\bx}$ and $\bomega$, we apply the Lorentz reciprocal theorem. Let $\hat{\bv},\hat{\sigma}$ be the velocity flow and the stress tensor of another solution of the Stokes equation which is regular over the domain $\mathbb{R}^3/\Omega$. The Lorentz reciprocal theorem then states that 
\begin{equation}
\int_{\partial \Omega} \bn \cdot \hat{\sigma} \cdot \bv = \int_{\partial \Omega} \bn \cdot \sigma \cdot \hat{\bv} \,.
\end{equation}
First, in order to get the squirmer's translation velocity, we choose $\hat{\bv},\hat{\sigma}$ to be the flow generated by a translation at velocity $\bU$ of the sphere $\Omega$ by an external force $\hat{\bF}$. The no-slip boundary condition then reads $\left. \hat{\bv} \right|_{\partial\Omega} = \hat{\bU}$. We therefore obtain
\begin{equation}
\hat{\bF} \cdot \dot{\bx} + \int_{\partial \Omega} \bn \cdot \hat{\sigma} \cdot \left(\bv_{s} + a \, \bomega \wedge \bn\right) = \bF \cdot \hat{\bU} \,.
\end{equation}
For a sphere of radius $a$, it leads to
\begin{equation}
\dot{\bx} = \frac{1}{6\pi\eta a}\bF - \frac{1}{4\pi a^2}\int_{\partial \Omega} \dd S \, \bv_{s}(\br,\bu) \,,
\end{equation}
independently of the angular velocity $\bomega$, since $\bn \cdot \hat{\sigma}$ is constant along the surface of the sphere. We then recover Eq.~\eqref{eq:translation}, in the absence of a background flow, with the self-propulsion speed
\begin{equation}
    v_0 = - \frac{1}{4\pi a^2}\int_{\partial \Omega} \dd S \, \bv_{s}(\br,\bu)\cdot\bu \,,
\end{equation}
and the mobility $\mu = 1/(6\pi\eta a)$. In order to obtain $\bomega$, we apply the Lorentz reciprocal theorem by considering $\hat{\bv},\hat{\sigma}$ to be the flow generated by a solid rotation at angular velocity $\hat{\bomega}$ of $\Omega$. On $\partial\Omega$, we have $\hat{\bv} = a \, \hat{\bomega} \wedge \bn$ and $\bn \cdot \hat{\sigma} = 3 \eta \, \hat{\bomega} \wedge \bn$, see \cite{stone1996propulsion}. We therefore obtain
\begin{equation}
    3 \eta \, \epsilon_{\alpha\beta\gamma}  \int_{\partial\Omega} \dd S \, n^\gamma \left(\dot{x}^\alpha + v_{s}^\alpha + a \epsilon_{\alpha\mu\nu}\omega^\mu n^\nu\right) = a \epsilon_{\alpha\beta\gamma}\int_{\partial\Omega} \dd S n^\rho \sigma^{\rho\alpha}n^\gamma \,,
\end{equation}
yielding
\begin{equation} \label{eq:rot_velocity}
    \bomega = \frac{1}{8\pi\eta a^3}\bGamma - \frac{3}{8\pi a^3}\int_{\partial\Omega}\dd S \,  \bn \wedge \bv_{s}(\br,\bu) \,.
\end{equation}
The equation of motion for the director $\bu$ then reads
\begin{equation}
    \dot{\bu} = \bomega \wedge \bu = \frac{1}{8\pi\eta a^3}\bGamma \wedge \bu \,,
\end{equation}
since the second term of Eq.~\eqref{eq:rot_velocity} points along $\bu$ by symmetry. We therefore recover the noiseless version of Eq.~\eqref{eq:rotation}, without the background flow, with the angular mobility $\mu_r = 1/(8\pi\eta a^3)$. In the presence of a background flow $\bar{\bv}$ the equations of motion can be found by considering the same Stokes equation imposing that at large distances the flow is equal to the background one, $\bv_\infty(\br) =  \bar{\bv}(\br)$. One can then obtain a formulation similar to Eqs.~\eqref{eq:stokes_app1}-\eqref{eq:div_app1}-\eqref{eq:bc_app1}, with a vanishing fluid flow at infinity, by considering $\hat{\bv}(\br) = \bv(\br) - \bv_\infty(\br)$. At the surface $\partial \Omega$, the corresponding boundary condition reads
\begin{equation}
    \left.\hat{\bv}\right|_{\partial\Omega}(\br) = \dot{\bx} + v_s(\br,\bu) + a \bomega \wedge \bn - \bar{\bv}(\br) \,.
\end{equation}
By denoting $\bx$ the position of the swimmer, one can then expand $\bar{\bv}(\br)$ around $\bar{\bv}(\bx)$ to first order in the radius $a$. Equations \eqref{eq:translation}-\eqref{eq:rotation} of the main text then follow from the application of the Lorentz reciprocal theorem as above. 

\section{Singularity of the angular dependence when $\epsilon_m=0$}\label{app:singular}
In this appendix, we consider the mode $m=0$ as an example. By incorrectly assuming that $\epsilon_0=0$, one obtains an equation for the angular dependence
\begin{equation}\label{eq:incorrect}
    \frac{1}{\sin\theta}\partial_\theta\left(\sin\theta \, \partial_\theta g_0 \right) + \lambda \sin\theta \partial_\theta g_0 + g_0 \left[2 + 4\lambda \cos\theta \right] = 0 \,.
\end{equation}
We now look for a perturbative solution in powers of the coupling constant $\lambda$ as $g_0(\theta) = g_0^{(0)}(\theta) + \lambda g_0^{(1)}(\theta) + \lambda^2 g_0^{(2)}(\theta) + \dots$. To leading order, we get
\begin{equation}
    g_0^{(0)}(\theta) = c^{(0)}_1 \cos\theta +c^{(0)}_2 \left[\frac{\cos\theta}{2}\log \left(\frac{1+\cos\theta}{1-\cos\theta}\right)-1\right] \,,
\end{equation}
with $c^{(0)}_1$ and $c^{(0)}_2$ two integration constants. We set $c^{(0)}_2=0$ to prevent divergence at $\cos\theta=\pm 1$ and choose $c^{(0)}_1=-1/4\pi$ to match known results for the Green function of the diffusion operator. Accordingly, to first order, we obtain
\begin{equation}
    g_0^{(1)}(\theta) = \frac{-10 \cos^2\theta+3 \cos\theta \log \left(\frac{1+\cos\theta}{1-\cos\theta}\right)}{32 \pi }+c^{(1)}_1 \cos\theta+c^{(1)}_2 \left[\frac{\cos\theta}{2} \log\left(\frac{1+\cos\theta}{1-\cos\theta}\right)-1\right] \,,
\end{equation}
with $c^{(1)}_1$ and $c^{(1)}_2$ two new integration constants. We then set $c^{(1)}_2 = - 3/16\pi$ for the solution to be well-behaved as $\cos\theta=\pm 1$. The integration constant $c^{(1)}_1$ is left undetermined so that
\begin{equation}
    g_0^{(1)}(\theta) = \frac{16 \pi  c^{(1)}_1 \cos\theta-5\cos^2\theta + 3}{16 \pi } \,.
\end{equation}
Using this we then evaluate $g_0^{(2)}(\theta)$ to find
\begin{equation}
\begin{split}
    g_0^{(2)}(\theta) & = \frac{60 \pi  c^{(1)}_1 \cos^2\theta - 18 \pi c^{(1)}_1 \cos\theta \log \left(\frac{1+\cos\theta}{1-\cos\theta}\right)-9 \cos^3\theta+2 \cos\theta \log (1-\cos^2\theta)}{48 \pi } \\ & + c^{(2)}_1 \cos\theta +c^{(2)}_2 \left[\frac{\cos\theta}{2}\log \left(\frac{1+\cos\theta}{1-\cos\theta}\right)-1\right] \,,
\end{split}
\end{equation}
with $c^{(2)}_1$ and $c^{(2)}_2$ two new integration constants. Hence, removing the log-divergence at both $\cos\theta=1$ and $\cos\theta=-1$ requires
\begin{equation}
    18 \pi  c^{(1)}_1-24 \pi  c^{(2)}_2+2 = 18 \pi  c^{(1)}_1-24 \pi  c^{(2)}_2-2 = 0 \,,
\end{equation}
which is impossible, so that no well-behaved solution can be found. This signals the emergence of a correction of the scaling dimension to order $O(\lambda^2)$.

\section{Renormalization group treatment of Eq.~\eqref{eq:density_far}}\label{app:RG}
In this appendix, we apply a perturbative renormalization group treatment to Eq.~\eqref{eq:density_far} to find the far-field decay of the density field. By linearity, this amounts to finding $K_\mu(\br)$, where 
\begin{equation}\label{eq:diff_adv}
    \Delta K_\mu(\br) - \lambda \, \partial_\alpha \left( v_\ell^\alpha(\br) K_\mu(\br) \right) = - \partial_\mu \delta(\br) \,,
\end{equation}
and where $\bv_\ell(\br) \equiv \bar{\bv}(\br)/(\lambda D)$ is such that 
\begin{equation}
    v^\alpha_\ell(\br) \simeq J^{\alpha\beta}(\br)\hat{p}^\beta \,,
\end{equation}
at large distances. For the sake of the renormalization group argument, the velocity field $\bv_\ell(\br)$ is explicitely built from a microscopic lengthscale $\ell$ as follows. First, we assume that the velocity field $v^\alpha_\ell(\br)$ can be expressed from a force density $q_\ell^\beta(\br)$, so that
\begin{equation}\label{eq:fluid_flow_ell}
     v_\ell^\alpha(\br) = \int \dd \br' J^{\alpha\beta}(\br-\br')q_\ell^\beta(\br') \,,
\end{equation}
with
\begin{equation}\label{eq:int_def_p}
    \int \dd \br \, q_\ell^\beta(\br) = \hat{p}^\beta \,.
\end{equation}
Then, we assume that the force density depends on a microscopic lengthscale $\ell$ through a scaling function $q^\beta$ according to
\begin{equation}
    q_\ell^\beta(\br) = \frac{1}{\ell^3} \, q^\beta\left(\frac{\br}{\ell}\right) \,. 
\end{equation}
We now look for a perturbative solution of Eq.~\eqref{eq:diff_adv} and study its behavior in the asymptotic regime where $\ell/\left|\br\right| \ll 1$. For any $\br$ finite, we obtain the solution up to order $O(\lambda^2)$ as
\begin{equation}\label{eq:integral_eq}
    \begin{split}
        K_\mu(\br) & =- \frac{1}{4\pi}\int \dd \br' \frac{1}{|\br-\br'|}\left(-\partial'_\mu \delta(\br') + \lambda \partial'_\alpha \left( v_\ell^\alpha(\br')K_\mu(\br')\right)\right) \\ & = - \frac{1}{4\pi}\frac{r^\mu}{r^3} + \frac{\lambda}{4\pi}\int \dd \br' v_\ell^\alpha(\br')K_\mu(\br') \frac{r^\alpha - r'^\alpha}{|\br-\br'|^3} \\ & = - \frac{1}{4\pi}\frac{r^\mu}{r^3} - \frac{1}{4\pi}\frac{\lambda}{4\pi}\int \dd \br' v_\ell^\alpha(\br') \frac{r'^\mu}{r'^3} \frac{r^\alpha - r'^\alpha}{|\br-\br'|^3} \\ & -\frac{1}{4\pi}\left(\frac{\lambda}{4\pi}\right)^2\int \dd \br' v_\ell^\alpha(\br') \int \dd \br'' v_\ell^\beta(\br'') \frac{r''^\mu}{r''^3} \frac{r'^\beta-r''^\beta}{|\br'-\br''|^3} \frac{r^\alpha - r'^\alpha}{|\br-\br'|^3}  + O(\lambda^3) \,.
    \end{split}
\end{equation}
In the following, we investigate the fate of this expansion in the far-field regime and use a renormalization group treatment to infer the anomalous scaling exponents.

\subsection{First order}\label{app:order1}

\noindent To first order in $\lambda$, we have
\begin{equation}
    K_\mu(\br) = - \frac{1}{4\pi}\frac{r^\mu}{r^3} - \frac{1}{4\pi}\frac{\lambda}{4\pi}\int \dd \br' v_\ell^\alpha(\br') \frac{r'^\mu}{r'^3} \frac{r^\alpha - r'^\alpha}{|\br-\br'|^3} + O(\lambda^2)
\end{equation}
We define
\begin{equation}\label{eq:def_I1}
\begin{split}
    I^\mu_1(\ell,\br) & \equiv \int \dd \br' v_\ell^\alpha(\br') \frac{r'^\mu}{r'^3} \frac{r^\alpha - r'^\alpha}{|\br-\br'|^3} \,, \\
    & = \int \dd \br' \int \dd \br'' J^{\alpha\beta}(\br'-\br'')\frac{1}{\ell^3} \, q^\beta\left(\frac{\br''}{\ell}\right) \frac{r'^\mu}{r'^3} \frac{r^\alpha - r'^\alpha}{|\br-\br'|^3}  \,.
\end{split}
\end{equation}
The latter can be brought to the scaling form of Eq.~\eqref{eq:scaling_form} by using dimensionless integration variables $\br'' \to \ell\br''$ and $\br' \to r\br'$ and using $J^{\alpha\beta}(\kappa \br) = \kappa^{-1}J^{\alpha\beta}(\br)$ for any positive number $\kappa > 0$,
\begin{equation}\label{eq:scaling_I1}
\begin{split}
    I^\mu_1(\ell,\br) & = \frac{1}{r^2} \int \dd \br' \int \dd \br'' J^{\alpha\beta}\left(\br' - \frac{\ell}{r} \br''\right)q^{\beta}(\br'')\frac{r'^\mu}{r'^3}\frac{\hat{r}^\alpha-r'^\alpha}{|\hat{\br}-\br'|^3} \\ & = \frac{1}{r^2}\hat{I}^\mu_1\left(\frac{\ell}{r},\hat{\br}\right)
\end{split}
\end{equation}
with
\begin{equation}
    \hat{I}^\mu_1\left(\epsilon,\hat{\br}\right) = \int \dd\br' \int \dd \br'' J^{\alpha\beta}(\br'-\epsilon \br'') q^\beta(\br'') \frac{r'^\mu}{r'^3}\frac{\hat{r}^\alpha-r'^\alpha}{|\hat{\br}-\br'|^3} \,.
\end{equation}
We now prove that the limit $\epsilon \to 0$ of the above integral exists. This amounts to showing that there is no anomalous scaling to first order in $\lambda$. To do so we first split the integral between a near-field and a far-field contribution
\begin{equation}
    \hat{I}^\mu_1\left(\epsilon,\hat{\br}\right) = J^\mu_1\left(\epsilon,\hat{\br}\right) + J^\mu_2\left(\epsilon,\hat{\br}\right) \,,
\end{equation}
with 
\begin{equation}
\begin{split}
    J^\mu_1\left(\epsilon,\hat{\br}\right) & = \int_0^{\sqrt{\epsilon}} \dd r' \int \dd \hat{\br}'  \int \dd \br'' J^{\alpha\beta}(\br'-\epsilon \br'') q^\beta(\br'') \hat{r}'^\mu \frac{\hat{r}^\alpha-r'^\alpha}{|\hat{\br}-\br'|^3} \,,
\end{split}
\end{equation}
and 
\begin{equation}
\begin{split}
    J^\mu_2\left(\epsilon,\hat{\br}\right) & = \int_{\sqrt{\epsilon}}^{+\infty} \dd r' \int \dd \hat{\br}'  \int \dd \br'' J^{\alpha\beta}(\br'-\epsilon \br'') q^\beta(\br'') \hat{r}'^\mu \frac{\hat{r}^\alpha-r'^\alpha}{|\hat{\br}-\br'|^3} \,.
\end{split}
\end{equation}
We can now evaluate the far-field $\epsilon \ll 1$ behavior of these integrals. Disregarding contributions vanishing when $\epsilon \to 0$, we obtain for the first one, 
\begin{equation}\label{eq:lim_J1}
\begin{split}
    J^\mu_1\left(\epsilon,\hat{\br}\right) & = \int_0^{1/\sqrt{\epsilon}} \dd r' \int \dd \hat{\br}'  \int \dd \br'' J^{\alpha\beta}(\br'-\br'') q^\beta(\br'') \hat{r}'^\mu\frac{\hat{r}^\alpha-\epsilon r'^\alpha}{|\hat{\br}-\epsilon \br'|^3} \simeq  \hat{J}_1^{\alpha\mu}  \hat{r}^\alpha \,,
\end{split}
\end{equation}
with the tensor
\begin{equation}
     \hat{J}_1^{\alpha\mu} \equiv \lim_{L \to \infty}  \int_0^L \dd r' \int \dd \hat{\br}' \int \dd \br'' J^{\alpha\beta}(\br'-\br'')q^\beta(\br'')\hat{r}'^\mu \,.
\end{equation}
We note that the above integral superficially seems logarithmically divergent as $L \to \infty$. Nonetheless, this divergence is prevented because the integral over the unit vector $\hat{\br}'$ vanishes at large distances. The tensor $\hat{J}_1^{\alpha\mu}$ is a non-universal correction, as it depends on the whole force distribution $q^\beta(\br)$. To leading order in the far field, the second integral becomes
\begin{equation}\label{eq:J2}
\begin{split}
    J^\mu_2\left(\epsilon,\hat{\br}\right) & \simeq \hat{J}_2^{\beta\mu}(\hat{\br}) \hat{p}^\beta
\end{split}
\end{equation}
with the tensor
\begin{equation}\label{eq:lim_J2}
    \hat{J}_2^{\beta\mu}(\hat{\br}) = \lim_{L\to 0}\int_{L}^{+\infty} \dd r' \int \dd \hat{\br}' J^{\alpha\beta}(\br') \, \hat{r}'^\mu \frac{\hat{r}^\alpha-r'^\alpha}{|\hat{\br}-\br'|^3} \,.
\end{equation}
Therefore, to leading order in the far field, and to order $O(\lambda)$ in the perturbation expansion, the solution reads
\begin{equation}
\begin{split}
    K_\mu(\br) & = - \frac{1}{4\pi r^2}\hat{r}^\mu - \frac{1}{4\pi r^2}\frac{\lambda}{4\pi}\left(\hat{J}_1^{\alpha\mu} \hat{r}^\alpha +  \hat{J}_2^{\alpha\mu}(\hat{\br}) \hat{p}^\alpha \right) + O(\lambda^2) \,, \\ & = -\frac{1}{4\pi r^2}\left(\delta^{\alpha\mu} + \frac{\lambda}{4\pi}\hat{J}_1^{\alpha\mu}\right)\left(\hat{r}^\alpha + \frac{\lambda}{4\pi} \hat{J}_2^{\beta\alpha}(\hat{\br}) \hat{p}^\beta\right) + O(\lambda^2) \,,
\end{split}
\end{equation}
and takes the form of a non-universal (tensorial) amplitude multiplied by a universal angular dependence. We now evaluate $\hat{J}_2^{\beta\alpha}(\hat{\br})$. Using isotropy, we can decompose
\begin{equation}
    \hat{J}_2^{\beta\alpha}(\hat{\br}) = A_1 \delta^{\beta\alpha}+ A_2 \hat{r}^\beta \hat{r}^\alpha \,,
\end{equation}
with
\begin{equation}
    A_1 = \frac{1}{2}\hat{J}_2^{\beta\alpha}(\hat{\br})\left(\delta^{\beta\alpha} - \hat{r}^\beta\hat{r}^\alpha\right) \,,
\end{equation}
and
\begin{equation}
    A_2 = - \frac{1}{2}\hat{J}_2^{\beta\alpha}(\hat{\br})\left(\delta^{\beta\alpha} - 3 \hat{r}^\beta\hat{r}^\alpha\right) \,.
\end{equation}
Then
\begin{equation}
    \begin{split}
        A_1 & = \frac{1}{2}\int_0^{+\infty}\dd r' \int \dd \hat{\br}'  \,\, \frac{\left(\delta^{\gamma\beta}+\hat{r}'^\gamma\hat{r}'^\beta\right)\hat{r}'^\alpha}{r'|\hat{\br}-\hat{\br}'|^3}\left(\hat{r}^\gamma-r' \hat{r}'^\gamma\right)\left(\delta^{\beta\alpha} - \hat{r}^\beta\hat{r}^\alpha\right) \\
        & = \frac{1}{2}\int_0^{+\infty}\dd r' \int \dd \hat{\br}' \frac{\left(\hat{r}^\gamma-r' \hat{r}'^\gamma\right)\hat{r}'^\alpha}{r'|\hat{\br}-\hat{\br}'|^3}\left(\delta^{\gamma\alpha}-\hat{r}^\gamma \hat{r}^\alpha+\hat{r}'^\gamma\hat{r}'^\alpha-\hat{r}^\alpha \hat{r}'^\gamma \left(\hat{\br}\cdot\hat{\br}'\right)\right) \\ & = \frac{1}{2}\int_0^{+\infty}\dd r' \int \dd \hat{\br}' \frac{\left(\hat{\br}\cdot\hat{\br}'\right)-\left(\hat{\br}\cdot\hat{\br}'\right)^3-2r'+2r' \left(\hat{\br}\cdot\hat{\br}'\right)^2}{r'|\hat{\br}-\hat{\br}'|^3} \\ & = \pi \int_0^{+\infty} \dd r'\int_{-1}^{-1} \dd w \, \frac{w - w^3 - 2r'+2r' w^2}{r'\left(1+r'^2-2 r' w\right)^{3/2}} \\ & = - 3 \pi \,.
    \end{split}
\end{equation}
Furthermore, 
\begin{equation}
    \begin{split}
        A_2 & = -\frac{1}{2}\int_0^{+\infty}\dd r' \int \dd \hat{\br}' \frac{\left(\delta^{\gamma\beta}+\hat{r}'^\gamma\hat{r}'^\beta\right)\hat{r}'^\alpha}{r'|\hat{\br}-\hat{\br}'|^3}\left(\hat{r}^\gamma-r' \hat{r}'^\gamma\right)\left(\delta^{\beta\alpha} - 3\hat{r}^\beta\hat{r}^\alpha\right) \\
        & = -\frac{1}{2}\int_0^{+\infty}\dd r' \int \dd \hat{\br}' \frac{\left(\hat{r}^\gamma-r' \hat{r}'^\gamma\right)\hat{r}'^\alpha}{r'|\hat{\br}-\hat{\br}'|^3}\left(\delta^{\gamma\alpha}-3\hat{r}^\gamma \hat{r}^\alpha+\hat{r}'^\gamma\hat{r}'^\alpha-3\hat{r}^\alpha \hat{r}'^\gamma \left(\hat{\br}\cdot\hat{\br}'\right)\right) \\ & = \frac{1}{2}\int_0^{+\infty}\dd r' \int \dd \hat{\br}' \frac{\left(\hat{\br}\cdot\hat{\br}'\right) + 3\left(\hat{\br}\cdot\hat{\br}'\right)^3 + 2r' - 6r' \left(\hat{\br}\cdot\hat{\br}'\right)^2}{r'|\hat{\br}-\hat{\br}'|^3} \\ & = \pi \int_0^{+\infty} \dd r' \int_{-1}^{-1} \dd w \, \frac{ w + 3 w^3 + 2r' - 6r' w^2}{r'\left(1+r'^2-2 r' w\right)^{3/2}} \\ & = 5 \pi \,.
    \end{split}
\end{equation}
Then, to first order in $O(\lambda)$, and to leading order in the far field, we have
\begin{equation}\label{eq:res_order1}
\begin{split}
    K_\mu(\br) & = -\frac{1}{4\pi r^2}\left(\delta^{\alpha\mu} + \frac{\lambda}{4\pi}\hat{J}_1^{\alpha\mu}\right)\left(\hat{r}^\alpha - \frac{3\lambda}{4} \hat{p}^\alpha + \frac{5\lambda}{4} \left(\hat{\bp} \cdot \hat{\br} \right) \hat{r}^\alpha\right) + O(\lambda^2) \,.
\end{split}
\end{equation}

\subsection{Second order}\label{sec:order2}

\noindent To second order, we need to compute the following integral 
\begin{equation}
    \begin{split}
        I^\mu_2(\ell,\br) & \equiv \int \dd \br' v_\ell^\alpha(\br') \int \dd \br'' v_\ell^\beta(\br'') \frac{r''^\mu}{r''^3} \frac{r'^\beta-r''^\beta}{|\br'-\br''|^3} \frac{r^\alpha - r'^\alpha}{|\br-\br'|^3} \,,
    \end{split}
\end{equation}
which enters Eq.~\eqref{eq:integral_eq}. Note that 
\begin{equation}
     I^\mu_2(\ell,\br) =  \int \dd \br' v_\ell^\alpha(\br') I^\mu_1(\ell,\br') \frac{r^\alpha - r'^\alpha}{|\br-\br'|^3} = \int \dd \br' v_\ell^\alpha(\br') \frac{1}{r'^2}\hat{I}^\mu_1\left(\frac{\ell}{r'},\hat{\br}'\right) \frac{r^\alpha - r'^\alpha}{|\br-\br'|^3} \,,
\end{equation}
as it appears from the definitions of $I^\mu_1(\ell,\br')$ and $\hat{I}^\mu_1(\ell/r',\hat{\br})$ in Eq.~\eqref{eq:def_I1} and Eq.~\eqref{eq:scaling_I1} respectively. This expression can then be brought to the scaling form of Eq.~\eqref{eq:scaling_form} using the same changes of variables as in Eq.~\eqref{eq:scaling_I1},
\begin{equation}
    \begin{split}
        I^\mu_2(\ell,\br) & = \int \dd \br' v_\ell^\alpha(\br') \frac{1}{r'^2}\hat{I}^\mu_1\left(\frac{\ell}{r'},\hat{\br}'\right) \frac{r^\alpha - r'^\alpha}{|\br-\br'|^3} \\ & = \int \dd \br' \int \dd \br'' J^{\alpha\beta}(\br'-\ell \br'') q^\beta(\br'')\frac{1}{r'^2}\hat{I}^\mu_1\left(\frac{\ell}{r'},\hat{\br}'\right) \frac{r^\alpha - r'^\alpha}{|\br-\br'|^3} \\ & = \frac{1}{r^2}\hat{I}^\mu_2\left(\frac{\ell}{r},\hat{\br}\right) \,,
    \end{split}
\end{equation}
with
\begin{equation}
    \hat{I}^\mu_2\left(\epsilon,\hat{\br}\right) = \int \dd \br' \int \dd \br'' J^{\alpha\beta}\left(\br'-\epsilon \br''\right) q^\beta(\br'')\frac{1}{r'^2}\hat{I}^\mu_1\left(\frac{\epsilon}{r'},\hat{\br}'\right) \frac{\hat{r}^\alpha - r'^\alpha}{|\hat{\br}-\br'|^3} \,.
\end{equation}
Again, we split the integral between a far field and a near-field contribution
\begin{equation}
    \hat{I}^\mu_2\left(\epsilon,\hat{\br}\right) = K^\mu_1(\epsilon,\hat{\br}) + K^\mu_2(\epsilon,\hat{\br}) \,,
\end{equation}
with
\begin{equation}\label{eq:def_K1}
    K^\mu_1(\epsilon,\hat{\br}) = \int_{0}^{\sqrt{\epsilon}} \dd r' \int \dd \hat{\br}' \int \dd \br'' J^{\alpha\beta}\left(\br'-\epsilon \br''\right) q^\beta(\br'')\hat{I}^\mu_1\left(\frac{\epsilon}{r'},\hat{\br}'\right) \frac{\hat{r}^\alpha - r'^\alpha}{|\hat{\br}-\br'|^3} \,,
\end{equation}
and 
\begin{equation}
    K^\mu_2(\epsilon,\hat{\br}) = \int_{\sqrt{\epsilon}}^{+\infty} \dd r' \int \dd \hat{\br}' \int \dd \br'' J^{\alpha\beta}\left(\br'-\epsilon \br''\right) q^\beta(\br'')\hat{I}^\mu_1\left(\frac{\epsilon}{r'},\hat{\br}'\right) \frac{\hat{r}^\alpha - r'^\alpha}{|\hat{\br}-\br'|^3} \,.
\end{equation}
We now investigate the behavior of both contributions when $\epsilon \ll 1$, neglecting vanishing corrections as $\epsilon \to 0$. For the second integral $K^\mu_2(\epsilon,\hat{\br})$, we have
\begin{equation}
\begin{split}
    K^\mu_2(\epsilon,\hat{\br}) & \simeq \hat{p}^\beta \int_{\sqrt{\epsilon}}^{+\infty} \dd r' \int \dd \hat{\br}' J^{\alpha\beta}\left(\br'\right) \hat{I}^\mu_1\left(0,\hat{\br}'\right) \frac{\hat{r}^\alpha - r'^\alpha}{|\hat{\br}-\br'|^3} \\ & \simeq \hat{p}^\beta \int_{\sqrt{\epsilon}}^{+\infty} \dd r' \int \dd \hat{\br}' J^{\alpha\beta}\left(\br'\right) \left[\hat{J}_1^{\gamma\mu} \hat{r}'^\gamma + \hat{J}_2^{\gamma\mu}(\hat{\br}')p^\gamma\right]\frac{\hat{r}^\alpha - r'^\alpha}{|\hat{\br}-\br'|^3} \\ & \simeq \hat{J}_1^{\gamma\mu}\hat{J}_2^{\beta\gamma}(\hat{\br}) \hat{p}^\beta + \hat{p}^\beta \hat{p}^\gamma \int_{\sqrt{\epsilon}}^{+\infty} \dd r' \int \dd \hat{\br}' J^{\alpha\beta}\left(\br'\right)  \hat{J}_2^{\gamma\mu}(\hat{\br}')\frac{\hat{r}^\alpha - r'^\alpha}{|\hat{\br}-\br'|^3} \,,
\end{split}
\end{equation}
where we used the far-field expression of $J_2(\epsilon,\hat{\br})$ in Eq.~\eqref{eq:J2} to get the first term on the right-hand side of the last equality. Crucially, because $J^{\alpha\beta}\left(\br\right) \sim r^{-1}$ and the angular integral does not vanish at short distances, the second term diverges logarithmically when $\epsilon \to 0$ and therefore contributes to the renormalization of the anomalous dimension to order $O(\lambda^2)$. We now focus on these diverging contributions which can be obtained by replacing the integrand by its small $r'$ behavior and using any finite number as an upper bound for the integral over $r'$, now represented as $\int_{\sqrt{\epsilon}} \dd r'$. As the logarithmically divergent part is insensitive to the upper bound, we get
\begin{equation}\label{eq:sing_K2}
\begin{split}
     \hat{p}^\beta \hat{p}^\gamma \int_{\sqrt{\epsilon}}^{+\infty} \dd r' \int \dd \hat{\br}' J^{\alpha\beta}\left(\br'\right)  \hat{J}_2^{\gamma\mu}(\hat{\br}')\frac{\hat{r}^\alpha - r'^\alpha}{|\hat{\br}-\br'|^3} & \sim \hat{r}^\alpha \hat{p}^\beta \hat{p}^\gamma \int_{\sqrt{\epsilon}} \dd r' \int \dd \hat{\br}' J^{\alpha\beta}\left(\br'\right)  \hat{J}_2^{\gamma\mu}(\hat{\br}') \,, \\ & \sim \hat{r}^\alpha \hat{p}^\beta \hat{p}^\gamma \int_{\sqrt{\epsilon}} \dd r' \frac{1}{r'} \int \dd \hat{\br}' \left(\delta^{\alpha\beta} + \hat{r}'^\alpha \hat{r}'^\beta\right)\left(-3\pi \delta^{\gamma\mu} + 5 \pi \hat{r}'^\gamma \hat{r}'^\mu \right) \,, \\ & \sim \ln \epsilon\left(\frac{10\pi^2}{3}\hat{p}^\mu\left(\hat{\br}\cdot\hat{\bp}\right) - \frac{2\pi^2}{3}\hat{r}^\mu\right) \,.
\end{split}
\end{equation}
To get the far-field angular dependence up to order $O(\lambda^2)$, it is further necessary to keep track of the terms in $K_2(\epsilon,\hat{\br})$ that remain finite as $\epsilon \to 0$. To do so, we introduce
\begin{equation}
\begin{split}
    Q^{\beta\gamma\mu}(\hat{\br}) & = \int_{\sqrt{\epsilon}}^{+\infty} \dd r' \int \dd \hat{\br}' J^{\alpha\beta}\left(\br'\right)  J_2^{\gamma\mu}(\hat{\br}')\frac{\hat{r}^\alpha - r'^\alpha}{|\hat{\br}-\br'|^3} \\ & = \int_{\sqrt{\epsilon}}^{+\infty} \dd r' \int \dd \hat{\br}' \, J^{\alpha\beta}(\br')  \left[-3\pi \delta^{\gamma\mu}+5\pi\hat{r}'^\gamma \hat{r}'^\mu\right]\frac{\hat{r}^\alpha - r'^\alpha}{|\hat{\br}-\br'|^3} \\ & = -3\pi \delta^{\gamma\mu}\int_{\sqrt{\epsilon}}^{+\infty} \dd r' \int \dd \hat{\br}' \, J^{\alpha\beta}(\br') \frac{\hat{r}^\alpha - r'^\alpha}{|\hat{\br}-\br'|^3} + 5 \pi \int_{\sqrt{\epsilon}}^{+\infty} \dd r' \int \dd \hat{\br}' \, J^{\alpha\beta}(\br') \frac{\hat{r}^\alpha - r'^\alpha}{|\hat{\br}-\br'|^3}\hat{r}'^\gamma \hat{r}'^\mu \\ & = - 3 \pi \delta^{\gamma\mu}Q_1^\beta(\hat{\br}) + 5 \pi Q_2^{\beta\gamma\mu}(\hat{\br}) \,,
\end{split}
\end{equation}
with
\begin{equation}
    Q_1^\beta(\hat{\br}) = \int_{\sqrt{\epsilon}}^{+\infty} \dd r' \int \dd \hat{\br}' \, J^{\alpha\beta}(\br') \frac{\hat{r}^\alpha - r'^\alpha}{|\hat{\br}-\br'|^3} \,,
\end{equation}
and 
\begin{equation}
    Q_2^{\beta\gamma\mu}(\hat{\br}) =  \int_{\sqrt{\epsilon}}^{+\infty} \dd r' \int \dd \hat{\br}' \, J^{\alpha\beta}(\br') \frac{\hat{r}^\alpha - r'^\alpha}{|\hat{\br}-\br'|^3}\hat{r}'^\gamma \hat{r}'^\mu \,.
\end{equation}
By symmetry, the vector $Q_1^\beta(\hat{\br})$ points along $\hat{r}^\beta$, meaning $Q_1^\beta(\hat{\br}) = Q_1 \hat{r}^\beta$ with
\begin{equation}
\begin{split}
    Q_1 & = \int_{\sqrt{\epsilon}}^{+\infty} \dd r' \int \dd \hat{\br}' \, J^{\alpha\gamma}(\br') \frac{\hat{r}^\alpha - r'^\alpha}{|\hat{\br}-\br'|^3}\hat{\br}^\gamma \\ & = 2\pi \int_{\sqrt{\epsilon}}^{+\infty} \dd r' \int_{-1}^{1} \dd w \frac{1 - 2 r' w + w^2}{r'\left(1+r'^2-2r' w\right)^{3/2}} \\ & = -\frac{8 \pi}{9} \left(1+3\ln \epsilon\right) \,. 
\end{split}
\end{equation}
The computation of $Q_2^{\beta\gamma\mu}(\hat{\br})$ is more tedious. We note that $Q_2^{\beta\gamma\mu}(\hat{\br})$ is symmetric under exchange of the indices $(\gamma,\mu)$. This leads to the decomposition 
\begin{equation}
    Q_2^{\beta\gamma\mu}(\hat{\br}) = B_1 \hat{r}^\beta \hat{r}^\gamma \hat{r}^\mu + B_2  \hat{r}^\beta \delta^{\gamma \mu} + B_3 \left( \hat{r}^\gamma \delta^{\beta\mu} + \hat{r}^\mu \delta^{\beta\gamma} \right) \,.
\end{equation}
We now evaluate $B_1$, $B_2$, and $B_3$ using the identities
\begin{equation}
\begin{split}
     & Q_2^{\beta\gamma\mu}(\hat{\br}) \hat{r}^\beta \hat{r}^\gamma \hat{r}^\mu = B_1 + B_2 + 2B_3 \,, \\
     & Q_2^{\beta\gamma\mu}(\hat{\br}) \hat{r}^\beta \delta^{\gamma\mu} = B_1 + 3 B_2 + 2 B_3 \,, \\
     & Q_2^{\beta\gamma\mu}(\hat{\br}) \hat{r}^\gamma \delta^{\beta\mu} = B_1 + B_2 + 4 B_3  \,,
\end{split}
\end{equation}
from which we obtain
\begin{equation}
\begin{split}
    B_1 & = \frac{1}{2} Q_2^{\beta\gamma\mu}(\hat{\br}) \left[5\hat{r}^\beta \hat{r}^\gamma \hat{r}^\mu -  \hat{r}^\beta \delta^{\gamma\mu} - 2 \hat{r}^\gamma \delta^{\beta\mu}\right] \,, \\
    B_2 & = \frac{1}{2} Q_2^{\beta\gamma\mu}(\hat{\br}) \left[-\hat{r}^\beta \hat{r}^\gamma \hat{r}^\mu + \hat{r}^\beta \delta^{\gamma\mu}\right] \,, \\
    B_3 & = \frac{1}{2} Q_2^{\beta\gamma\mu}(\hat{\br}) \left[-\hat{r}^\beta \hat{r}^\gamma \hat{r}^\mu + \hat{r}^\gamma \delta^{\beta\mu}\right] \,.
\end{split}
\end{equation}
Hence we have
\begin{equation}
\begin{split}
    B_1 & = \frac{1}{2}\int_{\sqrt{\epsilon}}^{+\infty} \dd r' \int \dd \hat{\br}' \, J^{\alpha\beta}(\br') \frac{\hat{r}^\alpha - r'^\alpha}{|\hat{\br}-\br'|^3}\hat{r}'^\gamma \hat{r}'^\mu \left[5\hat{r}^\beta \hat{r}^\gamma \hat{r}^\mu -  \hat{r}^\beta \delta^{\gamma\mu} - 2 \hat{r}^\gamma \delta^{\beta\mu}\right] \\
    & = \frac{1}{2}\int_{\sqrt{\epsilon}}^{+\infty} \dd r' \int \dd \hat{\br}' \, \frac{\hat{r}^\beta - \hat{r}'^\beta \left(2r' - \left(\hat{\br}\cdot \hat{\br}'\right)\right)}{r'|\hat{\br}-\br'|^3}\left(5 \hat{r}^\beta \left(\hat{\br}\cdot \hat{\br}'\right)^2 - \hat{r}^\beta - 2 \hat{r}'^\beta \left(\hat{\br}\cdot \hat{\br}'\right)\right) \\  & = \pi\int_{\sqrt{\epsilon}}^{+\infty} \dd r' \int_{-1}^{1} \dd w \frac{-1 -10 w^3 r' + 5 w^4 + 6 w r'}{r' \left(1+r'^2-2 w r'\right)^{3/2}} \\ & = \frac{12\pi}{5} \,.
\end{split}
\end{equation}
Similarly,
\begin{equation}
\begin{split}
    B_2 & = \frac{1}{2}\int_{\sqrt{\epsilon}}^{+\infty} \dd r' \int \dd \hat{\br}' \, J^{\alpha\beta}(\br') \frac{\hat{r}^\alpha - r'^\alpha}{|\hat{\br}-\br'|^3}\hat{r}'^\gamma \hat{r}'^\mu \left[-\hat{r}^\beta \hat{r}^\gamma \hat{r}^\mu + \hat{r}^\beta \delta^{\gamma\mu}\right] \\
    & = \frac{1}{2}\int_{\sqrt{\epsilon}}^{+\infty} \dd r' \int \dd \hat{\br}' \, \frac{\hat{r}^\beta - \hat{r}'^\beta \left(2r' - \left(\hat{\br}\cdot \hat{\br}'\right)\right)}{r'|\hat{\br}-\br'|^3}\left(- \hat{r}^\beta \left(\hat{\br}\cdot \hat{\br}'\right)^2 + \hat{r}^\beta\right) \\ & = \pi\int_{\sqrt{\epsilon}}^{+\infty} \dd r' \int_{-1}^{1} \dd w \frac{\left(1-w^2\right)\left(1-w\left(2r' - w\right)\right)}{r' \left(1+r'^2-2 w r'\right)^{3/2}} \\ & = - \frac{\pi}{5}\left(\frac{12}{5}+4\ln \epsilon\right) \,.
\end{split}
\end{equation}
Finally, we have
\begin{equation}
\begin{split}
    B_3 & = \frac{1}{2}\int_{\sqrt{\epsilon}}^{+\infty} \dd r' \int \dd \hat{\br}' \, J^{\alpha\beta}(\br') \frac{\hat{r}^\alpha - r'^\alpha}{|\hat{\br}-\br'|^3}\hat{r}'^\gamma \hat{r}'^\mu \left[-\hat{r}^\beta \hat{r}^\gamma \hat{r}^\mu + \hat{r}^\gamma \delta^{\beta\mu}\right] \\
    & = \frac{1}{2}\int_{\sqrt{\epsilon}}^{+\infty} \dd r' \int \dd \hat{\br}' \, \frac{\hat{r}^\beta - \hat{r}'^\beta \left(2r' - \left(\hat{\br}\cdot \hat{\br}'\right)\right)}{r'|\hat{\br}-\br'|^3}\left(-\hat{r}^\beta \left(\hat{\br}\cdot \hat{\br}'\right)^2 + \hat{r}'^\beta \left(\hat{\br}\cdot \hat{\br}'\right)\right) \\ & = \pi\int_{\sqrt{\epsilon}}^{+\infty} \dd r' \int_{-1}^{1} \dd w \frac{\left(2r'-w\right) \left(w^3-w\right)}{r' \left(1+r'^2-2 w r'\right)^{3/2}} \\ & = - \frac{\pi}{5}\left(\frac{208}{45} + \frac{2}{3}\ln \epsilon\right) \,.
\end{split}
\end{equation}
This leads to the following expression for $K^\mu_2(\epsilon,\hat{\br})$, where we keep all the terms that do not vanish as $\epsilon \to 0$
\begin{equation}
    \begin{split}
        K^\mu_2(\epsilon,\hat{\br}) & = \hat{p}^\beta \hat{J}_1^{\gamma\mu}\hat{J}_2^{\beta\gamma}(\hat{\br}) + \hat{p}^\beta \hat{p}^\gamma Q^{\beta\gamma\mu}(\hat{\br}) \\
        & = \hat{p}^\beta \hat{J}_1^{\gamma\mu}\hat{J}_2^{\beta\gamma}(\hat{\br}) + \hat{p}^\beta \hat{p}^\gamma \left[- 3 \pi \hat{r}^\beta \delta^{\gamma\mu}Q_1 + 5 \pi \left(B_1 \hat{r}^\beta \hat{r}^\gamma \hat{r}^\mu + B_2  \hat{r}^\beta \delta^{\gamma \mu} + B_3 \left( \hat{r}^\gamma \delta^{\beta\mu} + \hat{r}^\mu \delta^{\beta\gamma} \right)\right)\right] \\ & = \hat{p}^\beta \hat{J}_1^{\gamma\mu}\hat{J}_2^{\beta\gamma}(\hat{\br}) + \pi ^2 \left(\frac{10}{3} \log (\epsilon )-\frac{196}{45}\right) \hat{p}^\mu \left(\hat{\bp} \cdot \hat{\br}\right) + 12 \pi^2 \hat{r}^\mu  \left(\hat{\bp}\cdot \hat{\br}\right)^2 -\pi^2 \left(\frac{208}{45} + \frac{2}{3}\ln \epsilon\right)\hat{r}^\mu
    \end{split}
\end{equation}
Next, we expand $K^\mu_1(\epsilon,\hat{\br})$, defined in Eq.~\eqref{eq:def_K1}, when $\epsilon\ll 1$ and disregarding all the terms that vanish as $\epsilon\to 0$. First, we obtain
\begin{equation}\label{eq:integrand_to_replace}
\begin{split}
    K^\mu_1(\epsilon,\hat{\br}) & = \int_{0}^{\sqrt{\epsilon}} \dd r' \int \dd \hat{\br}' \int \dd \br'' J^{\alpha\beta}\left(\br'-\epsilon \br''\right) q^\beta(\br'')\hat{I}^\mu_1\left(\frac{\epsilon}{r'},\hat{\br}'\right) \frac{\hat{r}^\alpha - r'^\alpha}{|\hat{\br}-\br'|^3} \,, \\ 
    & = \int_{0}^{1/\sqrt{\epsilon}} \dd r' \int \dd \hat{\br}' \int \dd \br'' J^{\alpha\beta}\left(\br'-\br''\right) q^\beta(\br'')\hat{I}^\mu_1\left(\frac{1}{r'},\hat{\br}'\right) \frac{\hat{r}^\alpha - \epsilon r'^\alpha}{|\hat{\br}-\epsilon\br'|^3} \,, \\ & \simeq \hat{r}^\alpha \int_{0}^{1/\sqrt{\epsilon}} \dd r' \int \dd \hat{\br}' \int \dd \br'' J^{\alpha\beta}\left(\br'-\br''\right) q^\beta(\br'')\hat{I}^\mu_1\left(\frac{1}{r'},\hat{\br}'\right) \,.
\end{split}
\end{equation}
This last integral splits into two contributions, a finite non-universal contribution, denoted $\hat{K}_1^{\alpha\mu}$ in the following, and a universal logarithmically divergent contribution coming form the large distance behavior of the integral over $r'$. The latter can be obtained by replacing the integrand in Eq.~\eqref{eq:integrand_to_replace} by its large $r'$ leading order behavior (the integral over $\br''$ then gives $\hat{p}^\beta$ following Eq.~\eqref{eq:int_def_p}) and using the expression for $I_1^\mu(0,\hat{\br}')$ derived in Eqs.~\eqref{eq:lim_J1}-\eqref{eq:lim_J2}. In doing so, the lower bound in the $r'$ integral should be set to any strictly positive number, which we represent by $\int^{1/\sqrt{\epsilon}} \dd r'$. The logarithmically divergent contribution as $\epsilon \to 0$ is indeed insensitive to this bound. We therefore obtain
\begin{equation}\label{eq:singular_parts}
\begin{split}
     \hat{p}^\beta \int^{1/\sqrt{\epsilon}} \dd r' \int \dd \hat{\br}' J^{\alpha\beta}\left(\br'\right) \hat{I}^\mu_1\left(0,\hat{\br}'\right) & = \hat{p}^\beta \int^{1/\sqrt{\epsilon}} \dd r' \int \dd \hat{\br}' J^{\alpha\beta}\left(\br'\right) \left[ \hat{J}_1^{\gamma\mu} \hat{r}'^\gamma  +   \hat{J}_2^{\gamma\mu}(\hat{\br}') \hat{p}^\gamma\right] \,, \\ & \simeq \hat{p}^\beta \hat{p}^\gamma \int^{1/\sqrt{\epsilon}} \dd r' \int \dd \hat{\br}' J^{\alpha\beta}\left(\br'\right)  \hat{J}_2^{\gamma\mu}(\hat{\br}') \,, \\ & \simeq p^\beta p^\gamma \int_{\sqrt{\epsilon}} \dd r' \int \dd \hat{\br}' J^{\alpha\beta}\left(\br'\right)  \hat{J}_2^{\gamma\mu}(\hat{\br}')\,.
\end{split}
\end{equation}
Note that the term proportional to $\hat{J}_1^{\gamma\mu}$ that appears in the first line of Eq.~\eqref{eq:singular_parts} does not contribute to the singular part because the corresponding angular integral vanishes. Comparison with Eq.~\eqref{eq:sing_K2} then shows that the singular part of $K^\mu_1(\epsilon,\hat{\br})$ and that of $K^\mu_2(\epsilon,\hat{\br})$ are identical. We therefore obtain in the far field,
\begin{equation} \label{eq:I2_expr}
\begin{split}
    \hat{I}_2\left(\epsilon,\hat{\br}\right) & =  \left(\hat{K}_1^{\alpha\mu} - \delta^{\alpha\mu} \frac{4}{3}\pi^2 \ln \epsilon - \delta^{\alpha\mu} \frac{208}{45}\pi^2\right) \hat{r}^\alpha + \hat{p}^\beta \hat{J}_1^{\gamma\mu}\hat{J}_2^{\beta\gamma}(\hat{\br}) \\ & + \pi ^2 \left(\frac{20}{3} \log (\epsilon )-\frac{196}{45}\right) \hat{p}^\mu \left(\hat{\bp} \cdot \hat{\br}\right) + 12 \pi^2 \hat{r}^\mu  \left(\hat{\bp} \cdot \hat{\br}\right)^2  \,.
\end{split}
\end{equation}
Altogether, Eqs.~\eqref{eq:res_order1} and \eqref{eq:I2_expr} lead to the following expression for the expansion to second order $O(\lambda^2)$ of the solution of Eq.~\eqref{eq:diff_adv}, in the far field
\begin{equation}
    K_\mu(\br) = - \frac{1}{4\pi r^2}M^{\alpha\mu}\left(r\right)\left[\hat{r}^\alpha - \frac{\lambda}{4}\left(3 \hat{p}^\alpha - 5 \hat{r}^\alpha \left(\hat{\bp} \cdot \hat{\br}\right)\right) + \frac{3}{4} \lambda^2 \hat{r}^\alpha \left(\hat{\bp} \cdot \hat{\br}\right)^2 \right] + O(\lambda^3) \,,
\end{equation}
with the tensor
\begin{equation}
    M^{\alpha\mu}\left(r\right) = \delta^{\mu\alpha}\left(1-\frac{\lambda^2}{12}\ln \epsilon - \frac{13\lambda^2}{45}\right) + \frac{\lambda}{4\pi}\hat{J}_1^{\alpha\mu} + \left(\frac{\lambda}{4\pi}\right)^2 \hat{K}_1^{\alpha\mu} + \lambda^2\left(\frac{5}{12}\log (\epsilon )-\frac{49}{45}\right) \hat{p}^\alpha \hat{p}^\mu \,.
\end{equation}

\subsection{Renormalization group equations}
The far-field density decay is governed by Eq.~\eqref{eq:density_far} from which we get $\delta\rho(\br) = c^\mu K_\mu(\br)/D$. In the following, we show that, as in the treatment of the main text, we obtain two different anomalous dimensions depending on whether the polar obstacle has an axis of symmetry or not.

\subsubsection{Polar obstacle with an axis of symmetry}
If the obstacle has an axis of symmetry, the latter is necessarily along $\hat{\bp}$ so that $c^\mu = c \hat{p}^\mu$. Accordingly, by symmetry, we obtain $\hat{J}_1^{\alpha\mu}\hat{p}^\mu = j_1 \hat{p}^\alpha$ and $\hat{K}_1^{\alpha\mu}\hat{p}^\mu = k_1 \hat{p}^\alpha$ where $j_1$ and $k_1$ are constants which depend on near-field properties of the velocity field. We therefore get
\begin{equation}
\begin{split}
    c^\mu M^{\alpha\mu}(r) &= c \, \hat{p}^\alpha \left[1 + \frac{\lambda^2}{3}\lambda^2 \ln \epsilon - \frac{62}{45}\lambda^2 + \frac{\lambda}{4\pi}j_1 + \left(\frac{\lambda}{4\pi}\right)^2 k_1\right] \,, \\ & =  c \, \hat{p}^\alpha \left[1 - \frac{62}{45}\lambda^2 + \frac{\lambda}{4\pi}j_1 + \left(\frac{\lambda}{4\pi}\right)^2 k_1\right] \left(1 + \frac{\lambda^2}{3}\ln \epsilon\right) + O(\lambda^3) \,,
\end{split}
\end{equation}
which leads to the following expression for the density field
\begin{equation}\label{eq:renorm_sym}
    \delta\rho(\br) \propto \frac{m_{\parallel}(r)}{r^2}\left[\cos\theta - \frac{\lambda}{4}\left(3 - 5 \cos^2\theta\right) + \frac{3}{4} \lambda^2 \cos^3\theta \right] + O(\lambda^3)\,,
\end{equation}
with the function $m_{\parallel}(r)$ given by
\begin{equation} \label{eq:mparlog}
    m_{\parallel}(r) = 1 + \frac{\lambda^2}{3}\ln \frac{\ell}{r} \,.
\end{equation}
Note that equation \eqref{eq:renorm_sym} reproduces the angular dependence of Eq.~\eqref{eq:heur_res_bis}. This perturbative expansion is the first step of a renormalization group treatment done by introducing an arbitrary length scale $r'$ and writing
\begin{equation}
    m_{\parallel}(r) = m_{\parallel}(r')\left(1 + \frac{\lambda^2}{3}\ln \frac{r'}{r} \right) \,,
\end{equation}
which is valid up to order $O(\lambda^2)$. The renormalization group equation $\partial_{r'}m_{\parallel}(r) = 0$ therefore becomes
\begin{equation}\label{eq:RG_symm}
    \partial_{r'}m_{\parallel}(r') + \frac{\lambda^2}{3 r'}m_{\parallel}(r') = 0 \,,
\end{equation}
since the term scaling as $O\left(\lambda^2\partial_r'm_{\parallel}(r')\right)$ can be neglected to the considered order. Equation \eqref{eq:RG_symm} finally leads to
\begin{equation}
    \delta\rho(\br) \propto \frac{1}{r^{2 + \lambda^2/3}}\,,
\end{equation}
which reproduces the result of Eq.~\eqref{eq:heur_res}

\subsubsection{Obstacle with no axis of symmetry}
The situation is different when the obstacle doesn't have an axis of symmetry. In that case, we decompose $\boldsymbol{c} = c_{\parallel}\bp + \bc_{\perp}$ with $\bp \cdot \bc_{\perp}$ = 0. We isolate the logarithmically diverging contributions and split the different terms according to
\begin{equation}
\begin{split}
    \delta\rho(\br) = & -\frac{c_{\parallel}}{4\pi D r^2}\left(1 + \frac{\lambda^2}{3}\ln \frac{\ell}{r}\right)\left(\cos\theta - \frac{\lambda}{4}\left(3 - 5 \cos^2\theta\right) + \frac{3}{4} \lambda^2 \cos^3\theta \right) \\ & -\frac{1}{4\pi D r^2}\left(1 - \frac{\lambda^2}{12}\ln \frac{\ell}{r}\right)c^\alpha_{\perp}\left[\hat{r}^\alpha - \frac{\lambda}{4}\left(3 \hat{p}^\alpha - 5 \hat{r}^\alpha \left(\hat{\bp} \cdot \hat{\br}\right)\right) + \frac{3}{4} \lambda^2 \hat{r}^\alpha \left(\hat{\bp} \cdot \hat{\br}\right)^2 \right] \\ & - \frac{1}{4\pi D r^2} c^\mu \left[- \frac{13\lambda^2}{45}\delta^{\mu\alpha} + \frac{\lambda}{4\pi}\hat{J}_1^{\alpha\mu} + \left(\frac{\lambda}{4\pi}\right)^2 \hat{K}_1^{\alpha\mu} -\frac{49}{45} \lambda^2 \hat{p}^\alpha \hat{p}^\mu\right]\left[\hat{r}^\alpha - \frac{\lambda}{4}\left(3 \hat{p}^\alpha - 5 \hat{r}^\alpha \left(\hat{\bp} \cdot \hat{\br}\right)\right) + \frac{3}{4} \lambda^2 \hat{r}^\alpha \left(\hat{\bp} \cdot \hat{\br}\right)^2 \right]\,.
\end{split}
\end{equation}
Up to order $O(\lambda^2)$ we therefore obtain
\begin{equation}
\begin{split}
    \delta\rho(\br) = & -\frac{c_{\parallel}}{4\pi D r^2}\left(\frac{\ell}{r}\right)^{\lambda^2/3}\left(\cos\theta - \frac{\lambda}{4}\left(3 - 5 \cos^2\theta\right) + \frac{3}{4} \lambda^2 \cos^3\theta \right) \\ & -\frac{1}{4\pi D r^2}\left(\frac{\ell}{r}\right)^{-\lambda^2/12}c^\alpha_{\perp}\left[\hat{r}^\alpha - \frac{\lambda}{4}\left(3 \hat{p}^\alpha - 5 \hat{r}^\alpha \left(\hat{\bp} \cdot \hat{\br}\right)\right) + \frac{3}{4} \lambda^2 \hat{r}^\alpha \left(\hat{\bp} \cdot \hat{\br}\right)^2 \right] \\ & - \frac{1}{4\pi D r^2} c^\mu \left[- \frac{13\lambda^2}{45}\delta^{\mu\alpha} + \frac{\lambda}{4\pi}\hat{J}_1^{\alpha\mu} + \left(\frac{\lambda}{4\pi}\right)^2 \hat{K}_1^{\alpha\mu} -\frac{49}{45} \lambda^2 \hat{p}^\alpha \hat{p}^\mu\right]\left[\hat{r}^\alpha - \frac{\lambda}{4}\left(3 \hat{p}^\alpha - 5 \hat{r}^\alpha \left(\hat{\bp} \cdot \hat{\br}\right)\right) + \frac{3}{4} \lambda^2 \hat{r}^\alpha \left(\hat{\bp} \cdot \hat{\br}\right)^2 \right]\,.
\end{split}
\end{equation}
Hence the second line of the right-hand side dominates in the far field and we obtain
\begin{equation}
    \delta\rho(\br) \propto \frac{1}{r^{2-\lambda^2/12}}\cos\left(\phi + \phi_0\right)\sin\theta \left(1 + \frac{5}{4}\lambda \cos\theta + \frac{3}{4}\lambda^2 \cos^2\theta \right) \,,
\end{equation}
which reproduces Eqs.~\eqref{eq:heur_res_2} and \eqref{eq:heur_res_2bis} and where the phase $\phi_0$ is such that $\hat{\br}\cdot \bc_\perp = |\bc_\perp| \sin\theta \cos(\phi+\phi_0)$.

\bibliographystyle{unsrt}
\bibliography{biblio-activemf}

\end{document}